\documentclass[10pt,reqno]{amsart}

\def\showfigures{1}
\def\showtables{1}

\usepackage{amsmath,amssymb,amsfonts,dsfont}
\usepackage{cite,graphicx,xcolor,hyperref}

\renewcommand{\phi}{\varphi}
\newcommand{\LPT}{{L_2(\mathds{T})}}
\newcommand{\LPTs}{{L_2(\mathds{T}^*)}}
\newcommand{\LPTT}{{L_2(\mathds{T}^2)}}
\newcommand{\BasisT}{\{q(i,t)\}_{i=0}^\infty}
\newcommand{\BasisTT}{\{q(i,t) q(j,\tau)\}_{i,j=0}^\infty}
\newcommand{\trans}{{\scriptscriptstyle \mathrm{T}}}

\newcommand{\me}{\mathrm{e}}
\newcommand{\mi}{\mathrm{i}}
\renewcommand{\Re}{\mathop{\mathrm{Re}}}
\renewcommand{\Im}{\mathop{\mathrm{Im}}}

\newcommand{\bw}{{\scriptscriptstyle \mathrm{BW}}}
\newcommand{\lr}{{\scriptscriptstyle \mathrm{LR}}}
\newcommand{\ci}{{\scriptscriptstyle \mathrm{CI}}}
\newcommand{\cii}{{\scriptscriptstyle \mathrm{CII}}}

\author{K.\,A. Rybakov, E.\,D. Shermatov}

\title[Applying the spectral method for modeling linear filters]{Applying the Spectral Method \\ for Modeling Linear Filters: Butterworth, \\ Linkwitz--Riley, and Chebyshev Filters}

\begin{document}

\maketitle

\begin{center}

\vskip -3.5ex

Moscow Aviation Institute (National Research University);\\
125993, Moscow, Volokolamskoe Hwy, 4;\\
rkoffice@mail.ru, hdh.egor.shermatov@gmail.com
\end{center}

\vskip 2.5ex

\textbf{Abstract.} This paper proposes a new technique for computer modeling linear filters based on the spectral form of mathematical description of linear systems. It assumes the representation of input and output signals of the filter as orthogonal expansions, while filters themselves are described by two-dimensional non-stationary transfer functions. This technique allows one to model the output signal in continuous time, and it is successfully tested on the Butterworth, Linkwitz--Riley, and Chebyshev filters with different orders.

\vskip 0.5ex


\vskip 0.5ex


\thispagestyle{empty}
\makeatletter{\renewcommand*{\@makefnmark}{}
\footnotetext{Citation: Rybakov, K.A.; Shermatov, E.D. Applying the spectral method for modeling linear filters: Butterworth, Linkwitz–Riley, and Chebyshev filters. {\em Franklin Open} {\bf 2026}, {\em 14},~100508. \url{https://doi.org/10.1016/j.fraope.2026.100508}}\makeatother}

\section{Introduction}\label{secIntro}

The paper concerns the problem of signal processing by linear filters. A linear filter is a linear dynamic system that passes components of the input signal with frequencies from a certain range (passband) and suppresses components of the input signal with frequencies that do not belong to such a range (stopband)~\cite{Lam_82, Paa_03}.

These filters are designed to process analog signals, or signals in continuous time, and they are referred to as analog or electronic filters. They can be realized physically, e.g., as electronic circuits composed of resistors, inductors, and capacitors~\cite{Lam_82, LanSotThe_Sensors23}. The paper's primary focus is on low-pass filters, whose passband contains frequencies up to a predetermined frequency known as the cutoff frequency. High-pass filters are briefly discussed.

This field began to actively develop about a hundred years ago. The theory of linear filters is inextricably related to the research of Ronald Martin Foster~\cite{Fos_BSTJ24}, Wilhelm Cauer~\cite{Cau_31}, Stephen Butterworth~\cite{But_EWWE30}, Hendrik Wade Bode~\cite{Bode_JMP34}, Sidney Darlington~\cite{Dar_JMP39}, and others. The early stages of this theory are thoroughly described in reviews~\cite{Bel_PIRE62, Dar_TCS84}.

Despite the subsequent development of filtering methods, the results obtained by Cauer and Butterworth (the Butterworth and Chebyshev filters) are still in use. For example, they are used to process signals describing human physiological parameters~\cite{Bal_21, GraMeyHer_SR25}, in particular in electrocardiography and phonocardiography~\cite{Pal_19, ChoBanKimPal_22}, in monitoring both heart rate and saturation~\cite{WasTanHaf_JRESTI24}; are required to preprocess data in the noise reduction subsystem for an unmanned aerial vehicle~\cite{HuaHeYangLinOuJia_Drones24} and in a human activity recognition system~\cite{MaiBouGab_JAIHC23, MudAzmAla_SR25}; are used in electric motor or generator control systems~\cite{SungKangShim_JEET21}, for seismic wave analysis~\cite{SriPuhiSib_JS23} and monitoring of structures in motion caused by seismic vibrations~\cite{RodValCas_Infra24}; and are applied to noise compensation in autonomous navigation systems~\cite{AstZhiDem_20}, as well as part of the identity authentication mechanism in information systems~\cite{TianLiLiuLi_Sensors25}. Based on the Butterworth filters, the Linkwitz--Riley filters were proposed~\cite{Lin_JAES76}; they are traditionally applied in audio signal processing~\cite{HarVenChenMutDick_IEEE13, CecBruNobTerVal_JAES23}.

As a rule, the transition to discrete time is used for computer modeling linear filters~\cite{LutTosEva_01, OppSch_14, IngPro_17, GuGaoLiuMaoLia_IEEE23}. However, it is possible to propose a modeling technique that does not require time discretization (continuous time modeling is more natural for analog filters). This technique is based on the spectral form of mathematical description of linear systems~\cite{SolSemPeshNed_79, RybSot_TAC07, Ryb_Comp25}, which is oriented towards computer modeling from the moment of its appearance. The purpose of the paper is to describe and test the spectral method for modeling linear filters such as the Butterworth, Linkwitz--Riley, and Chebyshev filters. In this paper, the Linkwitz--Riley filters are considered to show how the spectral form is applied to describe a cascade of linear filters.

The spectral method is an approximate analytical method. It allows the output signal to be represented analytically as a continuous-time function. While this representation is theoretically exact, an approximate representation is used for computer modeling linear filters. The spectral method allows for the choice of a signal representation basis to ensure the desired properties. For example, orthogonal polynomials and trigonometric functions guarantee smoothness, or differentiability, as well as high approximation accuracy. Piecewise constant functions as a basis should produce results comparable to methods based on time discretization.

The proposed technique reduces filtering to a linear transform of expansion coefficients of both deterministic and random signals relative to the chosen basis, utilizing only algebraic operations. Thus, the spectral method combines the advantages of analytical methods and the simple implementation inherent in numerical methods. To the best of our knowledge, the spectral form of mathematical description has not been previously applied for computer modeling linear filters, in particular the Butterworth, Linkwitz--Riley, and Chebyshev filters.

Traditional methods based on time discretization have some drawbacks~\cite{OppSch_14}. For example, the impulse invariance method suffers from aliasing, which occurs if the discretization frequency is insufficient. The reason is that the information about signal is used only at discrete time moments, i.e., on a null set. In contrast, the spectral method uses the information about signal at a given time interval, and the information for discrete time moments of time is not significant. Another discretization method based on a bilinear transform suffers from frequency warping, which arises from the nonlinear nature of the frequency axis transform. The spectral method does not use any nonlinear transform of the frequency axis and, therefore, does not imply frequency warping.

The rest of the paper is structured as outlined below. In Section~\ref{secFilters}, four forms of mathematical description for linear filters are presented, and the problem statement and goal of the study are formulated. In Section~\ref{secSpectral}, the spectral form of mathematical description is presented in more detail. This section contains the necessary formulae for computer modeling linear filters in continuous time. The Butterworth, Linkwitz--Riley, and Chebyshev filters (Type I and Type II) are considered in Section~\ref{secFilterExamples}. For each filter family, the transfer function and two-dimensional non-stationary transfer function are presented. Section~\ref{secNumerical} is devoted to the testing methodology, and it also presents the results of the computational experiment. The paper is summarized in Section~\ref{secSpConcl}. Appendix contains an addition to the spectral form of mathematical description to take into account a phase delay.

\section{Description of Linear Filters and Problem Statement}\label{secFilters}

Here, we consider linear filters that can be described in any of the ways listed below. They are called forms of mathematical description of linear systems, and they differ in the mathematical framework that is used to describe input and output signals, as well as linear systems themselves~\cite{SolSemPeshNed_79}. The time is assumed to be continuous and bounded: $t \in \mathds{T} = [0,T]$.

In this section and below, we use standard notations and terms from real and complex analysis~\cite{KornKorn_00, Weg_12}. In particular, $\mathds{R}$ denotes the set of real numbers, $\mathds{C}$ represents the set of complex numbers (the complex plane); $\mi$ is the imaginary unit; $\Re z$ and $\Im z$ denote the real and imaginary parts of a number $z \in \mathds{C}$ which give the Cartesian coordinates of a point in the complex plane; $|z|$ and $\arg z$ are its absolute value and argument which determine the polar coordinates of a point on the complex plane; $\bar z$ is the complex conjugate of $z$. The sets $\{z \colon \Re z < 0\}$ and $\{z \colon \Im z > 0\}$ are the left and upper half-planes, respectively. In addition, some definitions and notations from the theory of function spaces are used, namely spaces of square-integrable functions $\LPT$ and $\LPTT$, as well as their bases~\cite{Bal_80}.

The expressions that relate input and output signals of a linear system using four forms of mathematical description may be different.

1.\;Description by linear differential equations:
\begin{equation}\label{eqODE}
  a_n x^{(n)}(t) + \ldots + a_1 x'(t) + a_0 x(t) = b_m g^{(m)}(t) + \ldots + b_1 g'(t) + b_0 g(t),
\end{equation}
where $g(t)$ and $x(t)$ are input and output signals, respectively, $a_0,a_1,\dots,a_n \in \mathds{R}$ and $b_0,b_1,\dots,b_m \in \mathds{R}$ are given values ($a_n \neq 0$ and $b_m \neq 0$), $n$ represents the filter order. The initial conditions are assumed to be zero.

2.\;Description by transition functions:
\begin{equation}\label{eqInOutIRF}
  x(t) = \int_0^t k(t,\tau) g(\tau) d\tau,
\end{equation}
where $k(t,\tau)$ is the impulse response function, i.e., the output signal of a linear system given the input signal as the Dirac delta function $\delta(t-\tau)$ under zero initial conditions. The impulse response function can be found by the equation~\eqref{eqODE}. Since coefficients of the left-hand and right-hand sides of the equation~\eqref{eqODE} are constant, the impulse response function can be considered as a function of one variable, namely $k(\eta)$ for $\eta = t - \tau$.

3.\;Description by integral transforms:
\begin{equation}\label{eqInOutLaplace}
  X(s) = H(s) G(s),
\end{equation}
where $G(s)$, $X(s)$, and $H(s)$ are defined by the Laplace transform ($\mathbb{L}$) of input and output signals, as well as the impulse response function as a function of one variable. In particular,
\[
  H = \mathbb{L} [k], \ \ \ H(s) = \int_0^{+\infty} k(\eta) \mathrm{e}^{-s\eta} d\eta.
\]

The function of a complex variable $H(s)$ is called the transfer function, it is expressed through coefficients $a_0,a_1,\dots,a_n$ and $b_0,b_1,\dots,b_m$:
\begin{equation}\label{eqDefH}
  H(s) = \frac{b_m s^m + \ldots + b_1 s + b_0}{a_n s^n + \ldots + a_1 s + a_0}.
\end{equation}

Along with the Laplace transform, the Fourier transform ($\mathbb{F}$) is used. The Fourier transform of the impulse response function as a function of one variable is expressed through the transfer function, namely $H(\mi \omega)$, where $\omega$ is a real variable having the meaning of frequency:
\[
  H = \mathbb{F} [k], \ \ \ H(\mi \omega) = \int_0^{+\infty} k(\eta) \mathrm{e}^{-\mi \omega \eta} d\eta.
\]
In the above formula, the physical realizability condition is taken into account, i.e., $k(\eta) = 0$ for $\eta \leqslant 0$. The function $H(\mi \omega)$ is called the frequency response function.

4.\;Description by spectral transforms:
\begin{equation}\label{eqInOutSp}
  X = W G,
\end{equation}
where $G$ and $X$ are non-stationary spectral characteristics of input and output signals, $W$ is the two-dimensional non-stationary transfer function of a linear system obtained by the spectral transform ($\mathbb{S}$).

The spectral form of mathematical description is infrequently used, and we will present it in more detail~\cite{SolSemPeshNed_79}. Let $\BasisT$ be an orthonormal basis of the space $\LPT$. The non-stationary spectral characteristic of a signal is the infinite column matrix formed by expansion coefficients of this signal relative to the basis $\BasisT$, i.e.,
\[
  G = \mathbb{S} [g], \ \ \ G_i = \int_\mathds{T} q(i,t) g(t) dt; \ \ \
  X = \mathbb{S} [x], \ \ \ X_i = \int_\mathds{T} q(i,t) x(t) dt,
\]
where $i = 0,1,2,\dots$

Based on the known non-stationary spectral characteristics, signals can be reconstructed as the following series:
\begin{equation}\label{eqSpInvert}
  g(t) = \sum\limits_{i=0}^\infty G_i q(i,t), \ \ \ x(t) = \sum\limits_{i=0}^\infty X_i q(i,t).
\end{equation}

Since $\BasisT$ is an orthonormal basis of $\LPT$, it follows that $\BasisTT$ is an orthonormal basis of $\LPTT$. The two-dimensional non-stationary transfer function of a linear system is the infinite matrix formed by expansion coefficients of the impulse response function as a function of two variables relative to the basis $\BasisTT$, i.e.,
\[
  W = \mathbb{S} [k], \ \ \ W_{ij} = \int_{\mathds{T}^2} k(t-\tau) q(i,t) q(j,\tau) dt d\tau,
\]
where $i,j = 0,1,2,\dots$

The matrix $W$ can be expressed through coefficients $a_0,a_1,\dots,a_n$ and $b_0,b_1,\dots,b_m$:
\begin{equation}\label{eqDefW}
  W = (a_n P^n + \ldots + a_1 P + a_0 E)^{-1} (b_m P^m + \ldots + b_1 P + b_0 E),
\end{equation}
where $P$ is the two-dimensional non-stationary transfer function of the derivative block, i.e., a linear system whose output signal is equal to the derivative of the input signal, and $E$ is the infinite identity matrix. The matrix $P^l$ can be considered as the two-dimensional non-stationary transfer function of the $l$th-order derivative block ($l = 0,1,\dots,n$), setting $P^0 = E$.

A linear filter can be interpreted as a linear operator defined on the signal space. In explicit form, this linear operator is specified by the formula~\eqref{eqInOutIRF}, and the impulse response function $k(t,\tau)$ defines its kernel. The two-dimensional non-stationary transfer function $W$ gives the spectral, or matrix, representation of such a linear operator relative to the signal space basis. In this paper, the impulse response function can be considered as a function of one variable, namely $k(\eta)$ for $\eta = t - \tau$, and it corresponds to the transfer function $H(s)$.

The problem is to describe linear filters, including the Butterworth, Linkwitz--Riley, and Chebyshev filters, in the spectral form. The main goal of such a description is to present a new technique for computer modeling linear filters. This technique is universal and can be applied to other linear filters. We choose the Butterworth, Linkwitz--Riley, and Chebyshev filters as examples for testing this technique.

The spectral form of mathematical description is presented in more detail in the next section. In particular, it provides relations for elements of the matrix $P$ for an arbitrary basis and for the basis that is subsequently used for numerical calculations.

\section{The Spectral Form of Mathematical Description}\label{secSpectral}

Before considering a more detailed presentation of the spectral form of mathematical description, we add some remarks to the part presented in the previous section.

1.\;The term ``non-stationary'' is used when describing signals and linear systems in the spectral form. In~\cite{SolSemPeshNed_79}, as well as in earlier and later publications on the spectral theory of control systems, the non-stationarity is understood in the following sense. Boundaries of the time segment, on which a linear system should be analyzed, change over time (both boundaries may change, or only one of them, usually the right boundary), i.e., the time segment is non-stationary. In this paper, the time segment is assumed to be fixed, or stationary.

2.\;Modern publications on the spectral form of mathematical description use different terminology. For example, in~\cite{BagMikPanRyb_Springer20, RybYus_IOP20}, the presentation is in terms of spectral characteristics of functions and linear operators. However, the results presented in this paper pertain to signal processing, and in this context, the classical terminology from~\cite{SolSemPeshNed_79} is more preferable.

3.\;Although when describing linear systems by integral and spectral transforms, signals and a linear system itself are represented differently (in the first case, as functions of a complex variable; in the second case, as infinite column matrices and infinite matrices), these forms exhibit similar properties. Both forms facilitate the transition from differential and integral relations to algebraic ones between input and output signals. Furthermore, for both forms, characteristics of a linear system are easily expressed through coefficients of the left-hand and right-hand sides of the equation~\eqref{eqODE}.

4.\;The formulae~\eqref{eqDefH} and~\eqref{eqDefW} have the same structure that enables to easily express the two-dimensional non-stationary transfer function by the transfer function and vice versa. In addition, the representation of the transfer function
\begin{equation}\label{eqDefHFactor}
  H(s) = \frac{b_m (s-\nu_1)(s-\nu_2) \ldots (s-\nu_m)}{a_n (s-s_1)(s-s_2) \ldots (s-s_n)},
\end{equation}
where $\nu_1,\nu_2,\dots,\nu_m$ and $s_1,s_2,\dots,s_n$ are zeros and poles of $H(s)$, corresponds to the representation of the two-dimensional non-stationary transfer function
\begin{equation}\label{eqDefWFactor}
  W = \frac{b_m}{a_n} \, (P - s_1 E)^{-1}(P - s_2 E)^{-1} \ldots (P - s_n E)^{-1} (P - \nu_1 E)(P - \nu_2 E) \ldots (P - \nu_m E).
\end{equation}

5.\;The domain for the spectral transform of functions of one variable is the space $\LPT$. However, the spectral transform can also be applied to generalized functions~\cite{SolSemPeshNed_79}, as well as random processes, including generalized random processes of the white noise type~\cite{Ryb_Comp25}. For the listed variants, the result of the spectral transform is the non-stationary spectral characteristic. The domain for the spectral transform of functions of two variables is also not limited to the space $\LPTT$, and generalized functions belong to such a domain. Four of the simplest linear systems are considered below, and only one of them corresponds to the impulse transfer function from $\LPTT$. Impulse transfer functions can be the generalized functions, and two-dimensional non-stationary transfer functions are defined for them.

6.\;When analyzing and synthesizing linear filters, frequency characteristics determined by the Fourier transform are used first. The spectral transform is associated with the generalized Fourier series, and arbitrary orthonormal bases can be used to represent signals. Below, trigonometric functions are used as a basis, but other bases can be used with the same success, e.g., orthogonal polynomials, piecewise constant functions, etc.

Next, we present the formulae that allow one to calculate elements of two-dimensional non-stationary transfer functions for four elementary blocks of control systems (four simplest linear systems): proportional, integral, and derivative blocks, as well as the pure time shift block~\cite{SolSemPeshNed_79}. We denote the corresponding matrices by $A,P^{-1},P$, and $S$.

The elements of the matrix $A$ (for the proportional block with gain $a(t)$) are calculated using the formula
\[
  A_{ij} = \int_\mathds{T} a(t) q(i,t) q(j,t) dt, \ \ \ i,j = 0,1,2,\dots
\]

For elements of matrices $P^{-1}$ and $P$ (for integral and derivative blocks), the following relations are satisfied:
\[
  P_{ij}^{-1} = \int_\mathds{T} q(i,t) \int_0^t q(j,\tau) d\tau dt, \ \ \
  P_{ij} = q(i,0) q(j,0) + \int_\mathds{T} q(i,t) \, \frac{dq(j,t)}{dt} \, dt, \ \ \ i,j = 0,1,2,\dots,
\]
and elements of the matrix $S$ (for the pure time shift block by $\tau$) are determined by the expression
\[
  S_{ij} = \int_\mathds{T} q(i,t) q(j,t+\tau) dt, \ \ \ i,j = 0,1,2,\dots,
\]
where $\tau < 0$ corresponds to a time delay, and $\tau > 0$ corresponds to a time advance. As a rule, in the latter formula, basis functions $q(i,t)$ are extended by zero: $q(i,t+\tau) = 0$ for $t+\tau \notin \mathds{T}$, $i = 0,1,2,\dots$ However, we redefine them differently (further, it is described in more detail). In the problem under consideration, this improves the quality of modeling linear filters in examples.

As a remark, we note that $A = a E$ in the simplest case with $a(t) = a = \mathrm{const}$, and in the general case, $A$ is a symmetric matrix. Matrices $P^{-1}$ and $P$ are mutually inverse, i.e., $P^{-1} P = P P^{-1} = E$. Obviously, $S = E$ for $\tau = 0$.

The formulae~\eqref{eqDefW} and~\eqref{eqDefWFactor} use two-dimensional non-stationary transfer functions of the derivative block and proportional blocks with constant gains $a_0,a_1,\dots,a_n$ and $b_0,b_1,\dots,b_m$. However, it is possible to rewrite them in an equivalent form by utilizing the two-dimensional non-stationary transfer function of the integral block~\cite{SolSemPeshNed_79}.

The two-dimensional non-stationary transfer function of the pure time shift block is not used in the formulae~\eqref{eqDefW} and~\eqref{eqDefWFactor}, but it will be necessary for numerical calculations later along with the two-dimensional non-stationary transfer function of the proportional block with gain $a(t) = \chi_{[0,T-\tau]}(t)$, the indicator of the set $[0,T-\tau]$. The latter of them can be applied in the optimal filtering problem, which can also be solved by the spectral method~\cite{Ryb_IOP21}.

The formulae for calculating elements of matrices $A,P^{-1},P$, and $S$ relative to cosines
\[
  q(i,t) = \left\{ \begin{array}{ll}
    \displaystyle \sqrt{\frac{1}{T}} & \text{for} ~ i = 0 \\ [-2.0ex] \\
    \displaystyle \sqrt{\frac{2}{T}} \, \cos \frac{i \pi t}{T} & \text{for} ~ i = 1,2,\dots,
  \end{array} \right.
\]
forming the orthonormal basis of $\LPT$, are offered below. In these formulae, the indices take the following values: $i,j = 1,2,\dots$ and $i \neq j$.

For matrices $P^{-1}$ and $P$ corresponding to integral and derivative blocks, they are given in~\cite{SolSemPeshNed_79}:
\[
  P_{00}^{-1} = \frac{T}{2}, \ \ \ P_{0i}^{-1} = -P_{i0}^{-1} = \frac{\sqrt{2} T [1-(-1)^i]}{i^2 \pi^2}, \ \ \ P_{ii}^{-1} = 0, \ \ \ P_{ij}^{-1} = -P_{ji}^{-1} = \frac{2 T [(-1)^{i+j} - 1]}{(i^2-j^2) \pi^2}
\]
and
\[
  P_{00} = \frac{1}{T}, \ \ \ P_{0i} = (-1)^i P_{i0} = \frac{(-1)^i \sqrt{2}}{T}, \ \ \ P_{ii} = \frac{2}{T}, \ \ \ P_{ij} = (-1)^{i+j} P_{ji} = \frac{2[i^2 - (-1)^{i+j} j^2]}{T(i^2 - j^2)}.
\]

For both the matrix $A$ corresponding to the proportional block with gain $a(t) = \chi_{[0,T-\tau]}(t)$ and the matrix $S$ corresponding to the pure time shift block, they are obtained by the simplest integration rules and trigonometric identities (a detailed derivation is not discussed here):
\begin{gather*}
  A_{00} = \frac{\tau}{T}, \ \ \ A_{0i} = A_{i0} = \frac{\sqrt{2}}{i \pi} \sin \frac{i \pi \tau}{T}, \ \ \ A_{ii} = \frac{\tau}{T} + \frac{1}{2 i \pi} \sin \frac{2 i \pi \tau}{T}, \\
  A_{ij} = A_{ji} = \frac{2}{(i^2 - j^2) \pi} \biggl( i \sin \frac{i \pi \tau}{T} \cos \frac{j \pi \tau}{T} - j \cos \frac{i \pi \tau}{T} \sin \frac{j \pi \tau}{T} \biggr)
\end{gather*}
and
\[
  S_{00} = 1, \ \ \ S_{0i} = \frac{\sqrt{2} [(-1)^i-1]}{i \pi} \sin \frac{i \pi \tau}{T}, \ \ \ S_{i0} = 0, \ \ \ S_{ii} = \cos \frac{i \pi \tau}{T}, \ \ \ S_{ij} = \frac{2j [1-(-1)^{i+j}]}{(i^2 - j^2) \pi} \sin \frac{j \pi \tau}{T}.
\]

Formally, the domain of basis functions is the segment $\mathds{T}$. However, basis functions can be defined beyond $\mathds{T}$ in a natural way, i.e., values of basis functions beyond $\mathds{T}$ are determined by the same formula as on $\mathds{T}$. Such a definition is used for calculating elements of the matrix $S$, and another method is discussed in Appendix.

For numerical calculations, the formulae for elements of non-stationary spectral characteristics of the simplest signals relative to cosines are also required. These formulae are given below, and the index for them takes the following values: $i = 1,2,\dots$

The elements of the column matrix $F$ corresponding to the signal $f(t) = \sin \omega t$ are
\[
  F_0 = \frac{1 - \cos T \omega}{\sqrt{T} \omega}, \ \ \ F_i = \frac{T \sqrt{2T} \omega [(-1)^i \cos T \omega - 1]}{i^2 \pi^2 - T^2 \omega^2} \ \ \ \text{for} \ \ \ i \pi \neq T \omega, \ \ \ F_i = 0 \ \ \ \text{for} \ \ \ i \pi = T \omega.
\]

For elements of the column matrix $F$ corresponding to the signal $f(t) = \cos \omega t$, we have
\[
  F_0 = \frac{\sin T \omega}{\sqrt{T} \omega}, \ \ \ F_i = \frac{T \sqrt{2T} (-1)^{i+1} \omega \sin T \omega}{i^2 \pi^2 - T^2 \omega^2} \ \ \ \text{for} \ \ \ i \pi \neq T \omega, \ \ \ F_i = \sqrt{\frac{T}{2}} \ \ \ \text{for} \ \ \ i \pi = T \omega,
\]
where $\omega > 0$.

The result of the spectral transform of standard Gaussian white noise $\eta(t)$ is the random column matrix $Q$ whose elements are independent Gaussian random variables with zero mean and unit variance~\cite{Ryb_Comp25, RybYus_IOP20}.

\section{The Butterworth, Linkwitz--Riley, and Chebyshev Filters}\label{secFilterExamples}

This section describes various low-pass filters. For each of them, we give the transfer function $H(s)$ for the cutoff frequency $\hat \Omega = 1$ and the corresponding two-dimensional non-stationary transfer function $W$ which is further considered as a function of the matrix $P$, i.e., $W = W(P)$. For an arbitrary cutoff frequency $\hat \Omega$, the transfer function and two-dimensional non-stationary transfer function are defined as follows:
\[
  \hat H(s) = H(s/\hat \Omega), \ \ \ \hat W(P) = W(\hat \Omega^{-1} P).
\]

Below, we will consider four families of low-pass filters (filters in each family differ in order):

(1) The Butterworth filters (BW),

(2) The Linkwitz--Riley filters (LR),

(3) The Chebyshev Type I filters (CI),

(4) The Chebyshev Type II filters (CII),

\noindent
where we give in parentheses the short notation used as a superscript for the transfer function, two-dimensional non-stationary transfer function, etc.

The dependence of numerical parameters, transfer function, and two-dimensional non-stationary transfer function on the filter order is not indicated for simplicity. It is assumed that the filter order $n$ can be uniquely identified by the given formulae.

As was stated in Section~\ref{secIntro}, this paper deals with low-pass filters but high-pass filters can also be described in the spectral form. It is no more difficult than the description of low-pass filters due to the same structure of the formulae~\eqref{eqDefH} and~\eqref{eqDefW} for the transfer function and two-dimensional non-stationary transfer function. For high-pass filters, we can write
\[
  \check H(s) = H(\hat \Omega/s), \ \ \ \check W(P) = W(\hat \Omega P^{-1}).
\]

Note that an ideal low-pass filter has the frequency response function $\hat H(\mi \omega)$, for which $|\hat H(\mi \omega)| = 1$ if $\omega \leqslant \hat \Omega$ and $|\hat H(\mi \omega)| = 0$ if $\omega > \hat \Omega$. For an ideal high-pass filter, the following property of the frequency response function $\check H(\mi \omega)$ is fulfilled: $|\check H(\mi \omega)| = 1$ if $\omega \geqslant \hat \Omega$ and $|\check H (\mi \omega)| = 0$ if $\omega < \hat \Omega$. It is impossible to implement filters with such properties, but it is possible to propose filters whose frequency response functions approximate the frequency response functions of ideal filters. The approximate filter should be described by the equation~\eqref{eqODE} or the transfer function~\eqref{eqDefH}. Filters can be physically implemented using such an approximation, and the above families of filters differ precisely in the approximation method.

\subsection{The Butterworth Filters}

The transfer function of the $n$th-order Butterworth filter is
\begin{equation}\label{eqTFBW1}
  H^\bw(s) = \frac{1}{(s - s_1^\bw) \dots (s - s_n^\bw)} = \biggl[ \, \prod\limits_{k=1}^n (s - s_k^\bw) \biggr]^{-1},
\end{equation}
where poles $s_k^\bw$ of the transfer function are roots of the equation $z^{2n} = -1$ (it has $2n$ roots, uniformly distributed on the unit circle centered at zero; they form complex conjugate pairs), belonging to the upper half-plane, with coefficient $\mi$:
\begin{equation}\label{eqDefBWPoles}
  s_k^\bw = \mi \, \me^{\mi \, \frac{-\pi + 2 \pi k}{2n}} = -A_k + \mi B_k, \ \ \ k = 1,\dots,n,
\end{equation}
where
\begin{equation}\label{eqDefAB}
  A_k = \sin \frac{-\pi + 2 \pi k}{2n}, \ \ \ B_k = \cos \frac{-\pi + 2 \pi k}{2n}.
\end{equation}

Recall that multiplying $z \in \mathds{C} \backslash \{0\}$ by $\mi$ corresponds to a counterclockwise rotation of the point $z$ about zero by angle $\pi/2$. Thus, all points $s_k$ belong to the left half-plane, and the corresponding transfer function defines an asymptotically stable linear dynamic system. In fact, poles of $H^\bw(s)$ are roots of the equation $(-\mi s)^{2n} = -1$, belonging to the left half-plane.

According to the formula~\eqref{eqDefWFactor}, the transfer function $H^\bw(s)$ corresponds to the two-dimensional non-stationary transfer function
\begin{equation}\label{eqNTFBW1}
  W^\bw(P) = \bigl( (P - s_1^\bw E) \dots (P - s_n^\bw E) \bigr)^{-1} = \biggl[ \, \prod\limits_{k=1}^n (P - s_k^\bw E) \biggr]^{-1}.
\end{equation}

If $n$ is even, then all poles form complex conjugate pairs $(s_1^\bw,s_n^\bw)$, $(s_2^\bw,s_{n-1}^\bw)$, \dots, $(s_r^\bw,s_{n-r+1}^\bw)$ for $r = \lfloor n/2 \rfloor$, where $\lfloor \, \cdot \, \rfloor$ denotes the integer part. If $n$ is odd, then $H^\bw(s)$ has the real pole $s_{r+1}^\bw = -1$ corresponding to the root $z = \mi$ of the equation $z^{2n} = -1$, and the remaining poles form complex conjugate pairs that are defined in the same way as for even $n$.

The domain coloring technique~\cite{Weg_12} can be used to visualize the characteristic polynomial of the Butterworth filter (the transfer function denominator). Here, a version of this technique based on the HLS color model is applied, i.e.,
\[
  \text{``Hue''} \propto \arg H^\bw(s), \ \ \ \text{``Lightness''} \propto \arctan |H^\bw(s)|, \ \ \ \text{``Saturation''} = 100\%.
\]

This technique assumes that the value of a function of a complex variable is coded by color; namely, hue and lightness correspond to argument and absolute value, while saturation takes the maximum value. It allows one to plot the graph of the function $f(z) \colon \mathds{C} \to \mathds{C}$ on a plane, which corresponds to the function $\tilde f \colon \mathds{R}^2 \to \mathds{R}^2$. Converting colors to grayscale preserves information about the absolute value and, consequently, about zeros of the function (zeros of the function corresponds to black dots on the graph). However, information about the argument will be lost.

\begin{figure}[ht]
  \centering
  \ifnum \showfigures = 1
  \includegraphics[scale = 1]{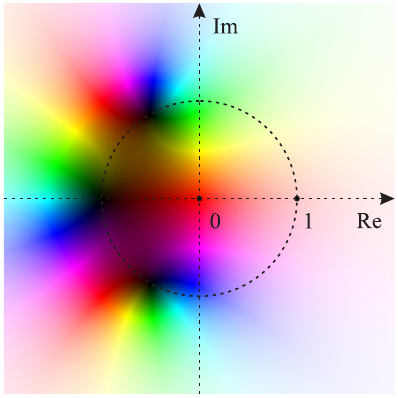}
  \fi
  \caption{The characteristic polynomial of the Butterworth filter ($n = 3$)}\label{picButterworthPoly}
\end{figure}

Figure~\ref{picButterworthPoly} shows a fragment of the complex plane $\{z \colon \Re z,\Im z \in [-2,2]\}$, the filter order is $n = 3$. Zeros of the characteristic polynomial (transfer function poles) correspond to three points on the unit circle: $-1$, $(-1 \pm \mi \sqrt {3})/2$.

The product of poles forming any complex conjugate pair (listed above) is equal to one, and their sum is equal to the doubled real part:
\[
  s_k^\bw = \bar s_{n-k+1}^\bw, \ \ \ s_k^\bw s_{n-k+1}^\bw = |s_k^\bw|^2 = A_k^2 + B_k^2 = 1, \ \ \ s_k^\bw + s_{n-k+1}^\bw = 2 \Re s_k^\bw = -2 A_k;
\]
therefore, $(s - s_k^\bw)(s - s_{n-k+1}^\bw) = s^2 + 2 A_k s + 1$. Consequently, the formula~\eqref{eqTFBW1} can be represented in the equivalent form
\begin{equation}\label{eqTFBW2}
  H^\bw(s) = \left\{
    \begin{array}{ll}
      \displaystyle \biggl[ \, \prod\limits_{k=1}^{n/2} (s^2 + 2 A_k s + 1) \biggr]^{-1} & \text{for even $n$} \\
      \displaystyle \biggl[ (s + 1) \prod\limits_{k=1}^{(n-1)/2} (s^2 + 2 A_k s + 1) \biggr]^{-1} & \text{for odd $n$},
    \end{array}
  \right.
\end{equation}
and the formula~\eqref{eqNTFBW1} can be rewritten as follows:
\begin{equation}\label{eqNTFBW2}
  W^\bw(P) = \left\{
    \begin{array}{ll}
      \displaystyle \biggl[ \, \prod\limits_{k=1}^{n/2} (P^2 + 2 A_k P + E) \biggr]^{-1} & \text{for even $n$} \\
      \displaystyle \biggl[ (P + E) \prod\limits_{k=1}^{(n-1)/2} (P^2 + 2 A_k P + E) \biggr]^{-1} & \text{for odd $n$}.
    \end{array}
  \right.
\end{equation}

\subsection{The Linkwitz--Riley Filters}

The Linkwitz--Riley filter is formed by cascading two Butterworth filters with the same order~\cite{Lin_JAES76}. This means that the order of such a filter is always even, and the transfer function of the $(2n)$th-order Linkwitz--Riley filter is represented as
\begin{equation}\label{eqTFLR}
  H^\lr(s) = \bigl( H^\bw(s) \bigr)^2,
\end{equation}
where $H^\bw(s)$ is the transfer function of the $n$th-order Butterworth filter (the Linkwitz--Riley filters are also called the Butterworth squared filters). Poles of these transfer functions coincide, and only the orders of poles differ. The transfer function $H^\lr(s)$ can be written in expanded form similar to the formulae~\eqref{eqTFBW1} and~\eqref{eqTFBW2}; however, it is sufficient to restrict ourselves to the formula~\eqref{eqTFLR}.

The two-dimensional non-stationary transfer function of the Linkwitz--Riley filter is obviously expressed as follows:
\begin{equation}\label{eqNTFLR}
  W^\lr(P) = \bigl( W^\bw(P) \bigr)^2;
\end{equation}
for it, expressions similar to the formulae~\eqref{eqNTFBW1} and~\eqref{eqNTFBW2} can also be proposed.

\begin{figure}[ht]
  \centering
  \ifnum \showfigures = 1
  \includegraphics[scale = 1]{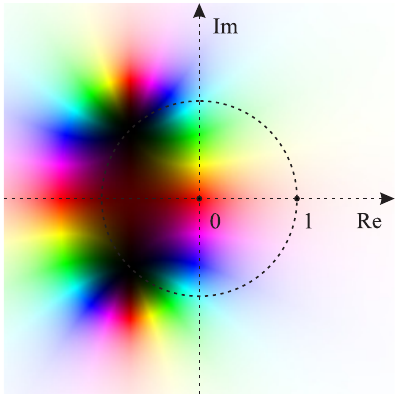}
  \fi
  \caption{The characteristic polynomial of the Linkwitz--Riley filter ($n = 4$)}\label{picLinkwitzRileyPoly}
\end{figure}

A visualization of the characteristic polynomial of the Linkwitz--Riley filter (the transfer function denominator) uses the above technique; the filter order is $n = 4$, see Figure~\ref{picLinkwitzRileyPoly}. Zeros of the characteristic polynomial (transfer function poles) correspond to two points on the unit circle: $(-1\pm \mi)/\sqrt{2}$.

\subsection{The Chebyshev Type I Filters}

To determine the transfer function of the $n$th-order Chebyshev Type I filter, the Chebyshev polynomial of the first kind with degree $n$ is used; it can be represented as follows~\cite{KornKorn_00}: $T_n(z) = \cos(n \arccos z)$. The transfer function of this filter is written in the form
\begin{equation}\label{eqTFCI1}
  H^\ci(s) = \frac{1}{\gamma(s - s_1^\ci) \dots (s - s_n^\ci)} = \frac{1}{\gamma} \biggl[ \, \prod\limits_{k=1}^n (s - s_k^\ci) \biggr]^{-1},
\end{equation}
where $\gamma = 2^{n-1} \varepsilon$ is the normalizing coefficient (the parameter $\varepsilon > 0$ is the passband ripple factor), and poles of the transfer function are roots of the equation $\varepsilon^2 T_n^2(z) = -1$ (it has $2n$ roots located on the ellipse with the center at zero and the semi-axes $\beta$ and $\alpha$ relative to the real and imaginary axes, respectively, the formulae for them are given below; they form complex conjugate pairs), belonging to the upper half-plane, with the coefficient $\mi$:
\begin{equation}\label{eqDefCIPoles}
  s_k^\ci = -\alpha A_k + \mi \beta B_k = \alpha \Re s_k^\bw + \mi \beta \Im s_k^\bw, \ \ \ k = 1,\dots,n,
\end{equation}
where values $A_k$ and $B_k$ are specified by the formula~\eqref{eqDefAB} ($s_k^\bw$ are poles of the transfer function of the $n$th-order Butterworth filter, see the formula~\eqref{eqDefBWPoles}), and the remaining constants are determined by expressions
\[
  \alpha = \sinh \lambda, \ \ \ \beta = \cosh \lambda, \ \ \ \lambda = \frac{1}{n} \mathop{\mathrm{arsh}} \frac{1}{\varepsilon}.
\]

Thus, poles of $H^\ci(s)$ are roots of the equation $\varepsilon^2 T_n^2(-\mi s) = -1$, belonging to the left half-plane. These poles can be expressed through poles of the transfer function of the Butterworth filter by the Joukowski transformation~\cite{Weg_12}. It is well known that the function
\[
  f(z) = \frac{1}{2} \biggl( z + \frac{1}{z} \biggr)
\]
transforms the circle with center at zero and radius $r$ into the ellipse with the same center and semi-axes $\beta = \cosh (\ln r)$ and $\alpha = \sinh (\ln r)$. In this case, $\ln r = \lambda$, so points $\me^\lambda z$ are located on the circle with center at zero and radius $r = \me^\lambda$ if $|z| = 1$. Additionally, taking into account the rotation of the ellipse about zero by angle $\pi/2$ (this is a consequence of the multiplication by $\mi$), we have
\[
  s_k^\ci = \frac{1}{2} \biggl( \me^\lambda s_k^\bw - \frac{1}{\me^\lambda s_k^\bw} \biggr), \ \ \ k = 1,\dots,n,
\]
where
\[
  \me^\lambda = \exp \biggl\{ \frac{1}{n} \mathop{\mathrm{arsh}} \frac{1}{\varepsilon} \biggr\} =
  \exp \biggl\{ \frac{1}{n} \ln \biggl( \frac{1}{\varepsilon} + \sqrt{1 + \frac{1}{\varepsilon^2}} \biggr) \biggr\} =
  \biggl\{ \frac{1}{\varepsilon} + \sqrt{1 + \frac{1}{\varepsilon^2}} \biggr\}^{1/n},
\]
if we utilize relations between exponential and hyperbolic functions.

This implies that poles of the transfer function are roots of the equation $z^{2n} = - (1/\varepsilon + \sqrt {1 + 1/\varepsilon^2})^2$ that belong to the upper half-plane, transformed by $f(z)$, and then multiplied by $\mi$.

According to the formula~\eqref{eqDefWFactor}, the transfer function $H^\ci(s)$ corresponds to the two-dimensional non-stationary transfer function
\begin{equation}\label{eqNTFCI1}
  W^\ci(P) = \frac{1}{\gamma} \bigl( (P - s_1^\ci E) \dots (P - s_n^\ci E) \bigr)^{-1} = \frac{1}{\gamma} \biggl[ \, \prod\limits_{k=1}^n (P - s_k^\ci E) \biggr]^{-1}.
\end{equation}

\begin{figure}[ht]
  \centering
  \ifnum \showfigures = 1
  \includegraphics[scale = 1]{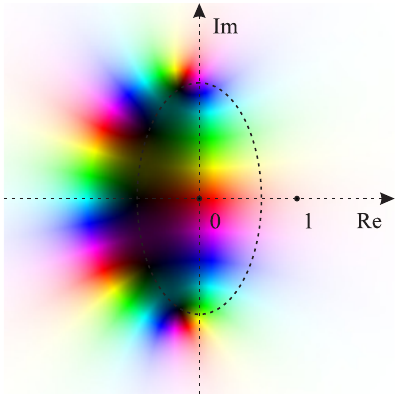}
  \fi
  \caption{The characteristic polynomial of the Chebyshev Type I filter ($n = 5$)}\label{picChebyshevIPoly}
\end{figure}

A visualization for the characteristic polynomial of the Chebyshev Type I filter (the transfer function denominator without the coefficient $\gamma$) is carried out in the same way as for the Butterworth filter; the filter order is $n = 5$, see Figure~\ref{picChebyshevIPoly}. Zeros of the characteristic polynomial (transfer function poles) correspond to five points; they lie on the ellipse that can be obtained by scaling the unit circle along the axes.

The analysis of poles for the Chebyshev Type I filter is similar to one for the Butterworth filter. The differences are that if $n$ is odd, then $H^\ci(s)$ has the real pole $s_{r+1}^\ci = -\alpha$ at $r = \lfloor n/2 \rfloor$, and also
\[
  s_k^\ci s_{n-k+1}^\ci = |s_k^\ci|^2 = (\alpha A_k)^2 + (\beta B_k)^2 \neq 1, \ \ \ s_k^\ci + s_{n-k+1}^\ci = 2 \Re s_k^\ci = -2 \alpha A_k.
\]

Introducing the notation $\rho_k^2 = (\alpha A_k)^2 + (\beta B_k)^2$, we obtain $(s - s_k^\ci)(s - s_{n-k+1}^\ci) = s^2 + 2 \alpha A_k s + \rho_k^2$. Therefore, the expression equivalent to the formula~\eqref{eqTFCI1} has the form
\begin{equation}\label{eqTFCI2}
  H^\ci(s) = \left\{
    \begin{array}{ll}
      \displaystyle \frac{1}{\gamma} \biggl[ \, \prod\limits_{k=1}^{n/2} (s^2 + 2 \alpha A_k s + \rho_k^2) \biggr]^{-1} & \text{for even $n$} \\
      \displaystyle \frac{1}{\gamma} \biggl[ (s + \alpha) \prod\limits_{k=1}^{(n-1)/2} (s^2 + 2 \alpha A_k s + \rho_k^2) \biggr]^{-1} & \text{for odd $n$},
    \end{array}
  \right.
\end{equation}
and we can use the following relation instead of the formula~\eqref{eqNTFCI1}:
\begin{equation}\label{eqNTFCI2}
  W^\ci(P) = \left\{
    \begin{array}{ll}
      \displaystyle \frac{1}{\gamma} \biggl[ \, \prod\limits_{k=1}^{n/2} (P^2 + 2 \alpha A_k P + \rho_k^2 E) \biggr]^{-1} & \text{for even $n$} \\
      \displaystyle \frac{1}{\gamma} \biggl[ (P + \alpha E) \prod\limits_{k=1}^{(n-1)/2} (P^2 + 2 \alpha A_k P + \rho_k^2 E) \biggr]^{-1} & \text{for odd $n$}.
    \end{array}
  \right.
\end{equation}

\subsection{The Chebyshev Type II Filters}

The transfer function of the $n$th-order Chebyshev Type II filter is specified using previously defined parameters (in this case, the parameter $\varepsilon > 0$  is the stopband ripple factor), namely
\begin{equation}\label{eqTFCII1}
  H^\cii(s) = \frac{\varkappa(s - \nu_1^\cii) \dots (s - \nu_n^\cii)}{(s - s_1^\cii) \dots (s - s_n^\cii)} = \varkappa \biggl[ \, \prod\limits_{k=1}^n (s - s_k^\cii) \biggr]^{-1} \prod\limits_{\begin{smallmatrix} k=1 \\ 2k-1 \neq n \end{smallmatrix}}^n (s - \nu_k^\cii),
\end{equation}
where
\begin{equation}\label{eqDefCIIPolesPre}
  \nu_k^\cii = \frac{\mi}{B_k} = \frac{\mi}{\Im s_k^\bw}, \ \ \ s_k^\cii = \frac{1}{s_k^\ci} = \frac{\bar s_k^\ci}{|s_k^\ci|^2}, \ \ \ k = 1,\dots,n,
\end{equation}
and $\varkappa = n^{n\,(\mathrm{mod}\,2)} \varepsilon$ is the normalizing coefficient ($\varkappa = \varepsilon$ for even $n$ and $\varkappa = n\varepsilon$ for odd $n$).

The product $(s - \nu_1^\cii) \dots (s - \nu_n^\cii)$ in the formula~\eqref{eqTFCII1} does not contain a factor corresponding to $B_k = 0$, and
\[
  \prod\limits_{\begin{smallmatrix} k=1 \\ 2k-1 \neq n \end{smallmatrix}}^n (s - s_k^\cii) = 1
\]
for $n = 1$ (a similar remark holds to other formulae in this subsection).

The transfer function zeros are roots of the equation $T_n(\mi/\nu) = 0$, which form complex conjugate pairs. They are expressed through zeros of the Chebyshev polynomial by the inversion and the coefficient~$\mi$. The transfer function poles can be defined as roots of the equation $\varepsilon^2 T_n^2(1/z) = -1$, belonging to the upper half-plane, with the coefficient $\mi$. Using relations~\eqref{eqDefBWPoles},~\eqref{eqDefCIPoles}, and~\eqref{eqDefCIIPolesPre}, we have
\begin{equation}\label{eqDefCIIPoles}
  s_k^\cii = -\frac{\alpha A_k}{\rho_k^2} - \mi \, \frac{\beta B_k}{\rho_k^2} = \frac{\alpha}{\rho_k^2} \Re s_k^\bw - \mi \, \frac{\beta}{\rho_k^2} \Im s_k^\bw, \ \ \ k = 1,\dots,n,
\end{equation}
i.e., poles of $H^\cii(s)$ are roots of the equation $\varepsilon^2 T_n^2 (-\mi/s) = -1$, belonging to the left half-plane. They can also be expressed in terms of poles of the transfer function of the Butterworth filter using the Joukowski transformation and inversion (the Chebyshev Type II filters are also called the inverse Chebyshev filters), i.e.,
\[
  s_k^\cii = 2 \biggl( \me^\lambda s_k^\bw - \frac{1}{\me^\lambda s_k^\bw} \biggr)^{-1}, \ \ \ k = 1,\dots,n.
\]

According to the formula~\eqref{eqDefWFactor}, the transfer function $H^\cii(s)$ corresponds to the two-dimensional non-stationary transfer function
\begin{equation}\label{eqNTFCII1}
  \begin{aligned}
    W^\cii(P) & = \varkappa \bigl( (P - s_1^\cii E) \dots (P - s_n^\cii E) \bigr)^{-1} (P - \nu_1^\cii E) \dots (P - \nu_n^\cii E) \\
    & = \varkappa \biggl[ \, \prod\limits_{k=1}^n (P - s_k^\cii E) \biggr]^{-1} \prod\limits_{\begin{smallmatrix} k=1 \\ 2k-1 \neq n \end{smallmatrix}}^n (P - \nu_k^\cii E).
  \end{aligned}
\end{equation}

\begin{figure}[ht]
  \centering
  \ifnum \showfigures = 1
  \includegraphics[scale = 1]{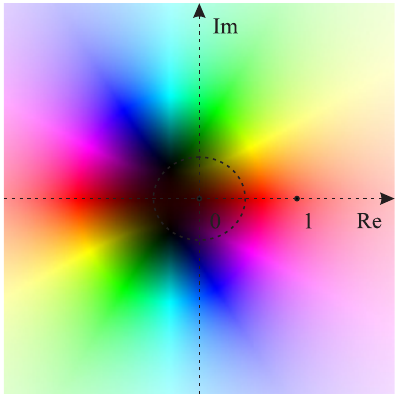}
  \fi
  \caption{The characteristic polynomial of the Chebyshev Type II filter ($n = 2$)}\label{picChebyshevIIPoly}
\end{figure}

A visualization for the characteristic polynomial of the Chebyshev Type II filter (the transfer function denominator) is shown in Figure~\ref{picChebyshevIIPoly}; the filter order is $n = 2$. Two points show zeros of the characteristic polynomial (transfer function poles).

For the Chebyshev Type II filter, the analysis of poles is carried out similarly:
\[
  s_k^\cii s_{n-k+1}^\cii = |s_k^\cii|^2 = \frac{1}{(\alpha A_k)^2 + (\beta B_k)^2} = \frac{1}{\rho_k^2}, \ \ \ s_k^\cii + s_{n-k+1}^\cii = 2 \Re s_k^\cii = -\frac{2 \alpha A_k}{\rho_k^2},
\]
and $(s - s_k^\cii)(s - s_{n-k+1}^\cii) = s^2 + 2 \alpha A_k s / \rho_k^2 + 1/\rho_k^2$. Additionally, $(s - \nu_k^\cii)(s - \nu_{n-k+1}^\cii) = s^2 + 1/|B_k|^2$. This yields the expression equivalent to the formula~\eqref{eqTFCII1}:
\begin{equation}\label{eqTFCII2}
  H^\cii(s) = \left\{
    \begin{array}{ll}
      \displaystyle \varkappa \biggl[ \, \prod\limits_{k=1}^{n/2} \biggl( s^2 + \frac{2 \alpha A_k}{\rho_k^2} \, s + \frac{1}{\rho_k^2} \biggr) \biggr]^{-1} ~ \prod\limits_{k=1}^{n/2} \biggl( s^2 + \frac{1}{|B_k|^2} \biggr) & \text{for even $n$}, \\
      \displaystyle \varkappa \biggl[ \biggl( s + \frac{1}{\alpha} \biggr) \prod\limits_{k=1}^{(n-1)/2} \biggl( s^2 + \frac{2 \alpha A_k}{\rho_k^2} \, s + \frac{1}{\rho_k^2} \biggr) \biggr]^{-1} ~ \prod\limits_{k=1}^{(n-1)/2} \biggl( s^2 + \frac{1}{|B_k|^2} \biggr) & \text{for odd $n$},
    \end{array}
  \right.
\end{equation}
and also the relation equivalent to the formula~\eqref{eqNTFCII1}:
\begin{equation}\label{eqNTFCII2}
  \begin{aligned}
    & W^\cii(P) \\
    & = \left\{
      \begin{array}{ll}
        \displaystyle \varkappa \biggl[ \, \prod\limits_{k=1}^{n/2} \biggl( P^2 + \frac{2 \alpha A_k}{\rho_k^2} \, P + \frac{1}{\rho_k^2} \, E \biggr) \biggr]^{-1} ~ \prod\limits_{k=1}^{n/2} \biggl( P^2 + \frac{1}{|B_k|^2} \, E \biggr) & \text{for even $n$}, \\
        \displaystyle \varkappa \biggl[ \biggl( P + \frac{1}{\alpha} \, E \biggr) \prod\limits_{k=1}^{(n-1)/2} \biggl( P^2 + \frac{2 \alpha A_k}{\rho_k^2} \, P + \frac{1}{\rho_k^2} \, E \biggr) \biggr]^{-1} ~ \prod\limits_{k=1}^{(n-1)/2} \biggl( P^2 + \frac{1}{|B_k|^2} \, E \biggr) & \text{for odd $n$}.
      \end{array}
    \right.
  \end{aligned}
\end{equation}

\section{Testing Methodology and Computational Experiment}\label{secNumerical}

This paper concerns well-known linear filters with properties that are described in detail in various design manuals~\cite{Lam_82, Paa_03, LutTosEva_01, IngPro_17}, so here testing does not mean an analysis of the amplitude and phase frequency characteristics of filters. Testing is related only to the spectral form, and it is based on a comparison of filter output signals with the original useful signal. This comparison is carried out in the spectral domain, and the transition to the time domain is implemented only for visualization of output signals.

We use
\[
  u(t) = \sin 10 \pi t
\]
as the useful signal and consider two noises. The first one is deterministic:
\[
  v(t) = \sigma (\sin 78 \pi t + \cos 95 \pi t + \sin 112 \pi t), \ \ \ \sigma = 0.2,
\]
and the second one is random:
\[
  v(t) = \sigma \eta(t), \ \ \ \sigma = 0.01,
\]
where $\eta(t)$ is standard Gaussian white noise.

The signal $g(t) = u(t) + v(t)$ is the filter input signal (the additive mixture of the useful signal and noise), and the filter output $x(t)$ is compared with the useful signal $u(t)$.

In order to obtain the spectral characteristic $G = \mathbb{S} [g]$, we can apply the linearity property of the spectral transform so that
\[
  G = U + V, \ \ \ U = \mathbb{S} [u], \ \ \ V = \mathbb{S} [v],
\]
where
\begin{gather*}
  V = \sigma (V_1 + V_2 + V_3) \ \ \ \text{for deterministic noise}, \ \ \ V = \sigma Q \ \ \ \text{for random noise}; \\
  V_1 = \mathbb{S} [v_1], ~ v_1(t) = \sin 78 \pi t; ~ V_2 = \mathbb{S} [v_2], ~ v_2(t) = \cos 95 \pi t; ~ V_3 = \mathbb{S} [v_3], ~ v_3(t) = \sin 112 \pi t; ~ Q = \mathbb{S} [\eta],
\end{gather*}
and the necessary formulae for elements of non-stationary spectral characteristics $U,V_1,V_2,V_3$, and $Q$ are given in Section~\ref{secSpectral}.

The spectral characteristic $X = \mathbb{S}[x]$ of the output signal is given by the expression~\eqref{eqInOutSp}, in which the two-dimensional non-stationary transfer function $W$ is determined by one of the formulae from Section~\ref{secFilterExamples}. The specific formula depends on the chosen filter family, filter order $n$, and cutoff frequency $\hat \Omega$. Further, we will assume that $\hat \Omega = 40 \pi$. The output signal in the time domain is described by the formula~\eqref{eqSpInvert}.

For linear filters, even if there is no noise, i.e., $v(t) = 0$, and the useful signal corresponds to the passband, then we should not expect the equality $x(t) = u(t)$ to be satisfied. This follows from properties of solutions to linear differential equations~\eqref{eqODE}. Therefore, the study of the error
\[
  \epsilon = \| x - u \|_\LPT
\]
in the problem under consideration is meaningless. For example, under the condition $v \in \LPT$, $\epsilon$ can be several times greater than $\|v\|_\LPT$ which can be taken as a priori error, even with good or excellent filter quality.

The output signal is characterized by a phase delay relative to the input signal. This phase delay can be exactly calculated~\cite{LutTosEva_01}:
\[
  \tau_\varphi = -\frac{\arg H(\mi \omega)}{\omega} \bigg|_{\omega = \Omega} \ \ \ (\Omega = 10 \pi),
\]
and it is advisable to compare signals $x^*(t) = x(t+\tau_\varphi)$ and $u(t)$ on $\mathds{T}^* = [0,T-\tau_\varphi]$. Therefore, we use the value
\[
  \mathcal{E} = \| x^* - u \|_\LPTs = \| \sqrt{\chi} \, (x^* - u) \|_\LPT,
\]
where $\chi_{[0,T-\tau_\varphi]}(t)$ is the indicator of $[0,T-\tau_\varphi]$.

The spectral form of mathematical description assumes that all the operations with signals are carried out in the spectral domain, i.e., with their expansion coefficients. Next, we show how this is implemented to take into account the phase delay and calculate the error $\mathcal{E}$.

Any admissible linear transform of signals corresponds to a linear transform of their non-stationary spectral characteristics in the spectral domain, and this linear transform is determined by a two-dimensional non-stationary transfer function. In particular, the time shift ($x \mapsto x^*$) corresponds to the expression $X^* = S X$, where $X^*= \mathbb{S} [x^*]$, and the matrix $S$ corresponds to the pure time shift block by $\tau_\varphi$. As a result, we obtain the expression for the output signal of a linear filter with phase delay compensation:
\begin{equation}\label{eqInOutSpDelay}
  X^* = S W G
\end{equation}
and
\begin{equation}\label{eqSpInvertDelay}
  x^*(t) = \sum\limits_{i=0}^\infty X_i^* q(i,t),
\end{equation}
where $X_i^*$ are elements of the column matrix $X^*$.

Further, we express the error $\mathcal{E}$. The right-hand side of the equality
\[
  \mathcal{E}^2 = \int_0^{T-\tau_\varphi} \bigl( x^*(t) - u(t) \bigr)^2 dt = \int_0^T \chi_{[0,T-\tau_\varphi]}(t) \bigl( x^*(t) - u(t) \bigr)^2 dt
\]
can be considered as the inner product for $\delta(t) = x^*(t) - u(t)$ and $\delta^*(t) = \chi_{[0,T-\tau_\varphi]}(t) \delta(t)$ in $\LPT$. The spectral characteristic $\Delta = \mathbb{S} [\delta]$ is expressed according to the linearity property of the spectral transform:
\[
  \Delta = X^* - G,
\]
and the multiplication ($\delta \mapsto \delta^*$) corresponds to the relation $\Delta^* = A \Delta$, where $\Delta^* = \mathbb{S} [\delta^*]$ and the matrix $A$ corresponds to the proportional block with gain $a(t) = \chi_{[0,T-\tau_\varphi]}(t)$. Additionally, taking into account that the spectral transform is the orthogonal transform that preserves the norm and inner product, we have
\[
  \mathcal{E}^2 = (\Delta^*)^\trans A \Delta \ \ \ \text{and} \ \ \ \mathcal{E} = \sqrt{(\Delta^*)^\trans A \Delta} = \sqrt{(X^* - U)^\trans A (X^* - U)},
\]
where $(\,\cdot\,)^\trans$ means the transpose of a matrix. It is useful to compare the error $\mathcal{E}$ with a priori error
\[
  \mathcal{E}_0 = \| v \|_\LPTs = \| \sqrt{\chi} \, v \|_\LPT,
\]
for which, by means of similar reasoning, we obtain the formula
\[
  \mathcal{E}_0 = \sqrt{V^\trans A V}.
\]

For more complex input signals, the group delay is usually applied instead of the phase delay,
\[
  \tau_g = -\frac{d\arg H(\mi \omega)}{d\omega},
\]
for the frequency $\omega$ that corresponds to the envelope of the useful signal~\cite{LutTosEva_01}.

The spectral form of mathematical description has an important aspect: it requires truncating all infinite column matrices and infinite matrices to finite sizes when used in computer modeling. Without such a truncation, the formulae provided earlier remain of theoretical interest only. The truncation means that all column matrices for describing signals should have size $L$, and matrices for describing linear systems should have size $L \times L$. When transiting from the spectral domain to the time domain, the following relations are used for the output signal (without phase delay compensation and with phase delay compensation, respectively):
\[
  x(t) \approx \sum\limits_{i=0}^{L-1} X_i q(i,t)\ \ \ \text{and} \ \ \ x^*(t) \approx \sum\limits_{i=0}^{L-1} X_i^* q(i,t).
\]

The calculation results for deterministic noise are presented below. For each of the filter families described in Section~\ref{secFilterExamples}, the errors $\mathcal{E}$ are given under the condition $\mathds{T} = [0,1]$, i.e., $T = 1$. For the Butterworth and Chebyshev filters, the orders are $n = 2,3,\dots,6$; for the Linkwitz--Riley filters, the orders are $n = 2,4,\dots,10$; for the Chebyshev filters, the ripple factor is $\varepsilon = 0.1$. The truncation orders defining sizes of matrices are $L = 128, 256, 512, 1024$.

The choice of truncation orders is motivated by several factors. The representation of the input signal as a partial sum of the Fourier series exhibits properties of a low-pass filter. To avoid this from affecting the results when testing linear filters, the truncation order should be chosen such that the additive mixture of the useful signal and noise is represented by the truncated non-stationary spectral characteristic without significant loss of accuracy. The truncation order $L = 128$ satisfies this condition.

Furthermore, $\log_2 128$ is an integer number. If only cosines are used as a basis, then this condition is not significant. However, such a choice is important for the Walsh and Haar functions (they are defined so that optimal truncation orders for them are $2^\mu$ for integer $\mu$). The truncation order $L = 128$ for cosines allows future comparisons of the presented results with those obtained by the Walsh and Haar functions.

The truncation orders $L = 256, 512, 1024$ are needed to analyze the dependence of the error on $L$ (doubling the truncation order is a common practice for the spectral method~\cite{Ryb_Comp25}, and it is similar to halving the integration step for numerical methods).

A priori error $\mathcal{E}_0$ depends not only on $L$ but also on $\tau_\varphi$. This implies that it depends on the chosen filter family and the filter order $n$. For example, for the third-order Butterworth filter, a priori error has values $\mathcal{E}_0 = 0.242112, 0.242318, 0.242319, 0.242319$ for $L = 128, 256, 512, 1024$, respectively. In order not to calculate a priori error for all considered filters, we restrict ourselves to the its estimate:
\[
  \mathcal{E}_0^+ = \sqrt{V^\trans V},
\]
depending only on $L$. Thus, according to calculations, we have $\mathcal{E}_0^+ = 0.244825, 0.245471, 0.245497, 0.245500$, and this is very close to $\mathcal{E}_0$.

The errors for the Butterworth filters are contained in Table~\ref{tabButterworthResult1}, for the Linkwitz--Riley filters in Table~\ref{tabLinkwitzRileyResult1}, and for the Chebyshev filters (Type I and Type II) in Tables~\ref{tabChebyshevIResult1} and~\ref{tabChebyshevIIResult1}, respectively.

\ifnum \showtables = 1

\begin{table}[ht]
\centering
\renewcommand{\arraystretch}{1.1}

\caption{The error $\mathcal{E}$ for the Butterworth filters (deterministic noise)}\label{tabButterworthResult1}
\begin{tabular}{ccccc}
  \hline
  $~~n~~$ & $L = 128$ & $L = 256$ & $L = 512$ & $L = 1024$ \\
  \hline
  2 & 0.047589 & 0.047764 & 0.047736 & 0.047737 \\
  3 & 0.024612 & 0.024531 & 0.024501 & 0.024493 \\
  4 & 0.015719 & 0.014764 & 0.014721 & 0.014679 \\
  5 & 0.011928 & 0.011391 & 0.011210 & 0.011187 \\
  6 & 0.010612 & 0.010224 & 0.010148 & 0.010130 \\
  \hline
\end{tabular}

\vskip 2ex

\caption{The error $\mathcal{E}$ for the Linkwitz--Riley filters (deterministic noise)}\label{tabLinkwitzRileyResult1}
\begin{tabular}{ccccc}
  \hline
  $~~n~~$ & $L = 128$ & $L = 256$ & $L = 512$ & $L = 1024$ \\
  \hline
  2 & 0.057454 & 0.057596 & 0.057607 & 0.057608 \\
  4 & 0.016257 & 0.014810 & 0.014751 & 0.014746 \\
  6 & 0.011089 & 0.010655 & 0.010572 & 0.010549 \\
  8 & 0.010621 & 0.010392 & 0.010256 & 0.010236 \\
 10 & 0.010860 & 0.010471 & 0.010366 & 0.010338 \\
  \hline
\end{tabular}

\vskip 2ex

\caption{The error $\mathcal{E}$ for the Chebyshev Type I filters (deterministic noise)}\label{tabChebyshevIResult1}
\begin{tabular}{ccccc}
  \hline
  $~~n~~$ & $L = 128$ & $L = 256$ & $L = 512$ & $L = 1024$ \\
  \hline
  2 & 0.171328 & 0.171823 & 0.171814 & 0.171808 \\
  3 & 0.065488 & 0.065305 & 0.065298 & 0.065301 \\
  4 & 0.021102 & 0.021020 & 0.020917 & 0.020907 \\
  5 & 0.012131 & 0.011623 & 0.011495 & 0.011465 \\
  6 & 0.010139 & 0.009888 & 0.009749 & 0.009709 \\
  \hline
\end{tabular}

\vskip 2ex

\caption{The error $\mathcal{E}$ for the Chebyshev Type II filters (deterministic noise)}\label{tabChebyshevIIResult1}
\begin{tabular}{ccccc}
  \hline
  $~~n~~$ & $L = 128$ & $L = 256$ & $L = 512$ & $L = 1024$ \\
  \hline
  2 & 0.040065 & 0.040035 & 0.039972 & 0.039971 \\
  3 & 0.026718 & 0.026629 & 0.026509 & 0.026503 \\
  4 & 0.014455 & 0.014213 & 0.014108 & 0.014072 \\
  5 & 0.022295 & 0.022069 & 0.022013 & 0.021994 \\
  6 & 0.022795 & 0.022501 & 0.022454 & 0.022444 \\
  \hline
\end{tabular}
\end{table}

\fi

Two components of the error $\mathcal{E}$ can be distinguished. The first one is the filter error, which is definitely present, since none of the filters under consideration are ideal. The second one is the spectral method error, associated with the truncation of all non-stationary spectral characteristics and two-dimensional non-stationary transfer functions. This component decreases with increasing truncation order $L$, which should be especially noticeable for smooth signals~\cite{CanHusQuaZan_06}. However, the errors presented in Tables~\ref{tabButterworthResult1}--\ref{tabChebyshevIIResult1} decrease rather slowly. These results indicate that the spectral method error is insignificant compared to the filter error.

The data in tables indicate the effectiveness of deterministic noise suppression, since the errors $\mathcal{E}$ are several times smaller than values $\mathcal{E}_0^+$ which estimate a priori error. In the considered example, an increase in $L$ is not required, since it does not lead to a noticeable improvement in the filter quality. The results from Tables~\ref{tabButterworthResult1}--\ref{tabChebyshevIIResult1} show that filter quality generally improves with increasing filter order.

The graphs of functions $u(t)$ (\textcolor{red}{red color}) and $g(t)$ (\textcolor{blue}{blue color}) are shown in Figure~\ref{picOriginalSignal1}.

\ifnum \showfigures = 1

\begin{figure}[ht]
  \centering
  \includegraphics[scale = 0.85]{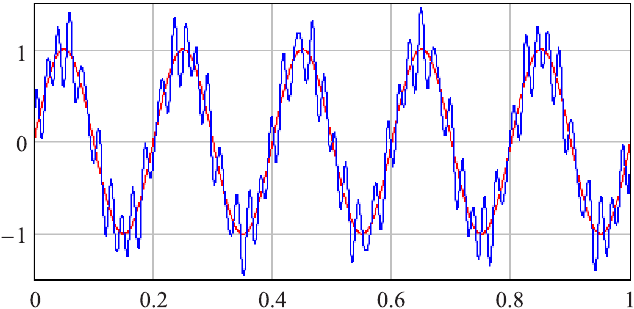}
  \caption{The graphs of the useful signal and the sum of useful signal and deterministic noise}\label{picOriginalSignal1}
\end{figure}

\fi

Selectively, i.e., for some $n$ and $L$, the filtering results are presented in Figures~\ref{picButterworthResult1}--\ref{picChebyshevIIResult1}: for $u(t)$ (\textcolor{red}{red color}) and $x^*(t)$ (\textcolor{blue}{blue color}). Noticeable discrepancies in the graphs of functions $g(t)$ and $x^*(t)$ occur only on $[T-\tau_\varphi,T]$, and this is quite expected.

\ifnum \showfigures = 1

\begin{figure}[p]
  \centering
  \includegraphics[scale = 0.85]{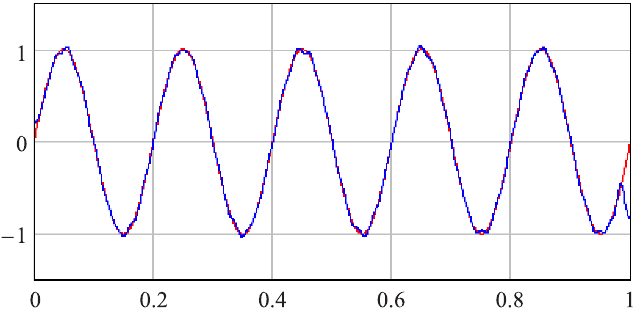}
  \caption{The graphs of the useful signal and the output signal of the Butterworth filter ($n = 3$, $L = 128$, deterministic noise)}\label{picButterworthResult1}
  \vskip 2ex
  \includegraphics[scale = 0.85]{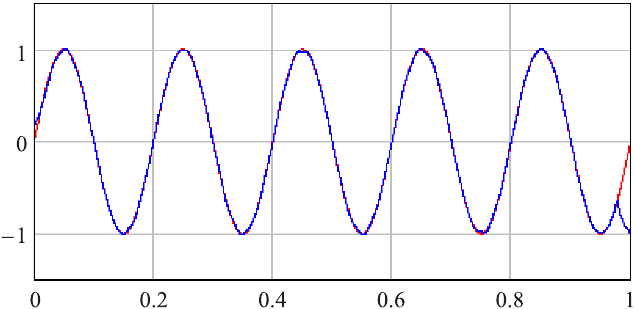}
  \caption{The graphs of the useful signal and the output signal of the Linkwitz--Riley filter ($n = 4$, $L = 256$, deterministic noise)}\label{picLinkwitzRileyResult1}
  \vskip 2ex
  \includegraphics[scale = 0.85]{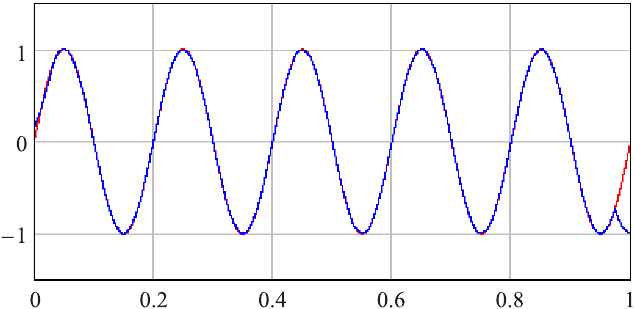}
  \caption{The graphs of the useful signal and the output signal of the Chebyshev Type~I filter ($n = 5$, $L = 512$, deterministic noise)}\label{picChebyshevIResult1}
  \vskip 2ex
  \includegraphics[scale = 0.85]{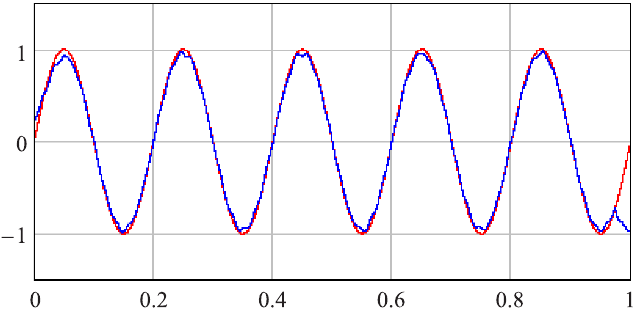}
  \caption{The graphs of the useful signal and the output signal of the Chebyshev Type~II filter ($n = 2$, $L = 1024$, deterministic noise)}\label{picChebyshevIIResult1}
\end{figure}

\fi

For random noise, the calculation technique is slightly more complicated. In this situation, input and output signals $g(t)$ and $x(t)$ are random processes, and their non-stationary spectral characteristics are random column matrices, so the error $\mathcal{E}$ is a random variable. Let $\mathcal{E}_j$ represent the realization of the random variable $\mathcal{E}$ that corresponds to realizations $g_j(t)$ and $x_j(t)$ of input and output signals, $j = 1,\dots,M$, where $M = 10^4$ is the number of realizations.

We use estimates of the mathematical expectation and standard deviation of the random variable $\mathcal{E}$ (the standard deviation is expressed through a biased estimate of the variance, but an unbiased estimate can be used instead):
\[
  \bar{\mathcal{E}} = \frac{1}{M} \sum\limits_{j=1}^M \mathcal{E}_j, \ \ \ \sigma_\mathcal{E} = \sqrt{\frac{1}{M} \sum\limits_{j=1}^M (\mathcal{E}_j - \bar{\mathcal{E}})^2}.
\]

For each of the filter families described in Section~\ref{secFilterExamples}, estimates of the mathematical expectation $\bar{\mathcal{E}}$ and the standard deviation $\sigma_\mathcal{E}$ are obtained. These values are given in Table~\ref{tabButterworthResult2} for the Butterworth filters, in Table~\ref{tabLinkwitzRileyResult2} for the Linkwitz--Riley filters, and in Tables~\ref{tabChebyshevIResult2} and~\ref{tabChebyshevIIResult2} for the Chebyshev filters (Type I and Type II, respectively). The data in these tables are presented in the following format: $\bar{\mathcal{E}}~(\sigma_\mathcal{E})$. The value $T$, filter orders $n$, and ripple factor $\varepsilon$ for the Chebyshev filters are the same as for deterministic noise. The truncation orders defining sizes of matrices are $L = 128, 256, 512$.

Here, $\mathcal{E}_0^+$ is also a random variable, and we can similarly find estimates of the mathematical expectation $\bar {\mathcal{E}}_0^+$ and the standard deviation $\sigma_{\mathcal{E}_0^+}$ for it:
\[
  \bar{\mathcal{E}}_0^+ = \frac{1}{M} \sum\limits_{j=1}^M (\mathcal{E}_0^+)_j, \ \ \ \sigma_{\mathcal{E}_0^+} = \sqrt{\frac{1}{M} \sum\limits_{j=1}^M \bigl( (\mathcal{E}_0^+)_j - \bar{\mathcal{E}}_0^+ \bigr)^2},
\]
where $(\mathcal{E}_0^+)_j$ is the realization of the random variable $\mathcal{E}_0^+$ that corresponds to the realization $v_j(t)$ of random noise (the same number of realizations $M$ is used). Thus, according to calculations, we have $\bar{\mathcal{E}}_0^+ = 0.112872, 0.159906, 0.226173$ and $\sigma_{\mathcal{E}_0^+} = 0.007051, 0.007093, 0.007090$ for $L = 128, 256, 512$, respectively. The increase in $\bar{\mathcal{E}}_0^+$ with increasing $L$ is due to the fact that standard Gaussian white noise is used as random noise, so $\bar{\mathcal{E}}_0^+ \to \infty$ as $L \to \infty$.

\ifnum \showtables = 1

\begin{table}[ht]
\centering
\renewcommand{\arraystretch}{1.1}

\caption{The error $\mathcal{E}$ for the Butterworth filters (random noise)}\label{tabButterworthResult2}
\begin{tabular}{cccc}
  \hline
  $~~n~~$ & $L = 128$ & $L = 256$ & $L = 512$ \\
  \hline
  2 & 0.065750~(0.006140) & 0.066069~(0.006140) & 0.065991~(0.006073) \\
  3 & 0.064114~(0.006371) & 0.063839~(0.006325) & 0.064141~(0.006360) \\
  4 & 0.063366~(0.006548) & 0.063248~(0.006611) & 0.063151~(0.006596) \\
  5 & 0.062816~(0.006658) & 0.062779~(0.006667) & 0.062742~(0.006667) \\
  6 & 0.062612~(0.006829) & 0.062378~(0.006607) & 0.062498~(0.006765) \\
  \hline
\end{tabular}

\vskip 2ex

\caption{The error $\mathcal{E}$ for the Linkwitz--Riley filters (random noise)}\label{tabLinkwitzRileyResult2}
\begin{tabular}{cccc}
  \hline
  $~~n~~$ & $L = 128$ & $L = 256$ & $L = 512$ \\
  \hline
  2 & 0.069214~(0.007119) & 0.069267~(0.007161) & 0.069370~(0.007141) \\
  4 & 0.057452~(0.006231) & 0.056937~(0.006261) & 0.056956~(0.006295) \\
  6 & 0.058282~(0.006528) & 0.057997~(0.006490) & 0.058098~(0.006584) \\
  8 & 0.058779~(0.006649) & 0.058706~(0.006690) & 0.058582~(0.006683) \\
 10 & 0.059029~(0.006784) & 0.058999~(0.006719) & 0.059055~(0.006689) \\
  \hline
\end{tabular}

\vskip 2ex

\caption{The error $\mathcal{E}$ for the Chebyshev Type I filters (random noise)}\label{tabChebyshevIResult2}
\begin{tabular}{cccc}
  \hline
  $~~n~~$ & $L = 128$ & $L = 256$ & $L = 512$ \\
  \hline
  2 & 0.096526~(0.006469) & 0.100908~(0.006239) & 0.101290~(0.006236) \\
  3 & 0.079350~(0.006596) & 0.079449~(0.006545) & 0.079384~(0.006530) \\
  4 & 0.071808~(0.006691) & 0.071758~(0.006694) & 0.071702~(0.006697) \\
  5 & 0.068183~(0.006861) & 0.068015~(0.006759) & 0.068038~(0.006843) \\
  6 & 0.066051~(0.006820) & 0.065990~(0.006856) & 0.065947~(0.006895) \\
  \hline
\end{tabular}

\vskip 2ex

\caption{The error $\mathcal{E}$ for the Chebyshev Type II filters (random noise)}\label{tabChebyshevIIResult2}
\begin{tabular}{cccc}
  \hline
  $~~n~~$ & $L = 128$ & $L = 256$ & $L = 512$ \\
  \hline
  2 & 0.055214~(0.007209) & 0.056249~(0.007106) & 0.058224~(0.006821) \\
  3 & 0.052184~(0.006383) & 0.052560~(0.006359) & 0.052775~(0.006335) \\
  4 & 0.056234~(0.006720) & 0.056485~(0.006577) & 0.058290~(0.006415) \\
  5 & 0.058212~(0.006699) & 0.059000~(0.006695) & 0.059625~(0.006630) \\
  6 & 0.060050~(0.006831) & 0.059860~(0.006811) & 0.061285~(0.006688) \\
  \hline
\end{tabular}
\end{table}

\fi

The analysis of errors with statistical characteristics from Tables~\ref{tabButterworthResult2}--\ref{tabChebyshevIIResult2} enables the formulation of the same conclusion: the spectral method error is insignificant compared to the filter error. The filters are less effective here, but this is clear since the noise contains components with frequencies from the passband. These factors mean that both the truncation order and the filter order have no noticeable effect on the results. However, filters under random noise are also effective because the estimate of the mathematical expectation $\bar{\mathcal{E}}$ is less than the estimate of the mathematical expectation $\bar{\mathcal{E}}_0^+$.

The graphs of functions $u(t)$ (\textcolor{red}{red color}) and $g(t)$ (\textcolor{blue}{blue color}) are depicted in Figure~\ref{picOriginalSignal2}, where a single realization of $v(t)$ is taken.

\ifnum \showfigures = 1

\begin{figure}[ht]
  \centering
  \includegraphics[scale = 0.85]{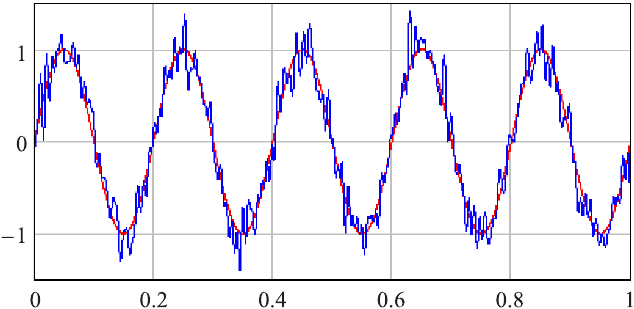}
  \caption{The graphs of the useful signal and the sum of useful signal and random noise}\label{picOriginalSignal2}
\end{figure}

\fi

Figures~\ref{picButterworthResult2}--\ref{picChebyshevIIResult2} show the results in graphical form for some values $n$ and $L = 256$: for $u(t)$ (\textcolor{red}{red color}) and $x^*(t)$ (\textcolor{blue}{blue color}); the graphs are shown for a single realization of $v(t)$.

\ifnum \showfigures = 1

\begin{figure}[p]
  \centering
  \includegraphics[scale = 0.85]{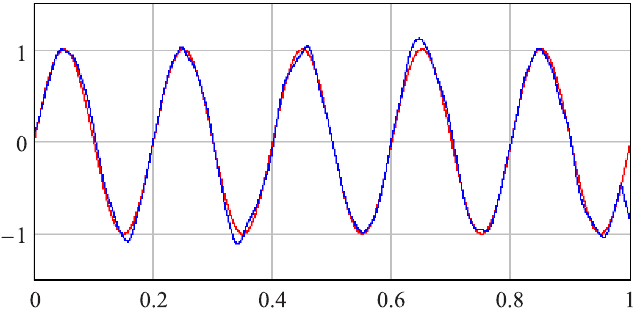}
  \caption{The graphs of the useful signal and the output signal of the Butterworth filter ($n = 3$, $L = 256$, random noise)}\label{picButterworthResult2}
  \vskip 2ex
  \includegraphics[scale = 0.85]{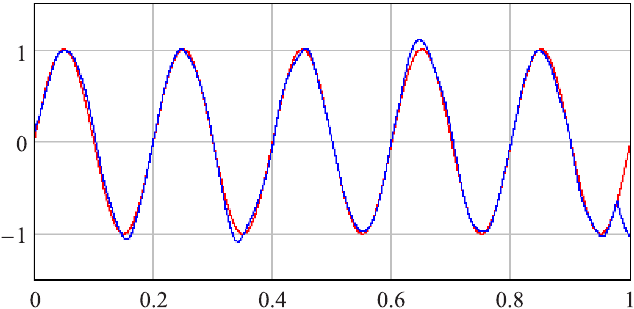}
  \caption{The graphs of the useful signal and the output signal of the Linkwitz--Riley filter ($n = 4$, $L = 256$, random noise)}\label{picLinkwitzRileyResult2}
  \vskip 2ex
  \includegraphics[scale = 0.85]{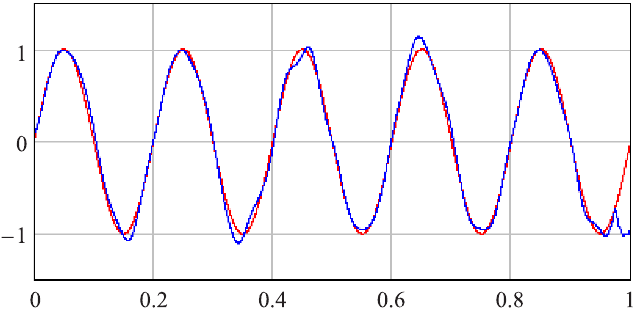}
  \caption{The graphs of the useful signal and the output signal of the Chebyshev Type~I filter ($n = 5$, $L = 256$, random noise)}\label{picChebyshevIResult2}
  \vskip 2ex
  \includegraphics[scale = 0.85]{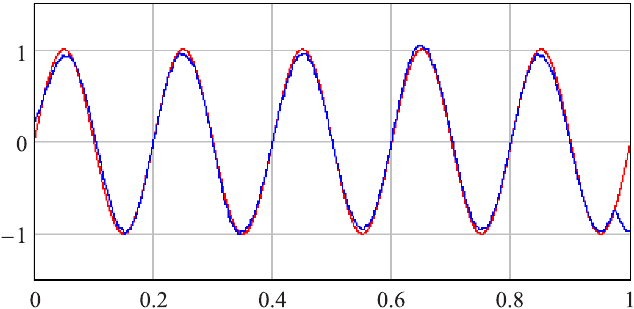}
  \caption{The graphs of the useful signal and the output signal of the Chebyshev Type~II filter ($n = 2$, $L = 256$, random noise)}\label{picChebyshevIIResult2}
\end{figure}

\fi

The computational complexity of finding the two-dimensional non-stationary transfer function for a linear filter can be estimated as $O(L^3)$. This follows from properties of matrix multiplication and inversion, and this estimate holds when using standard algorithms. When using fast algorithms, the computational complexity is lower. For example, if the Strassen algorithm is applied, then the computational complexity can be estimated as $O(L^{\log_2 7})$. The computational effort required to obtain the output signal, provided the known two-dimensional non-stationary transfer function, is proportional to the truncation order $L$.

\section{Conclusions}\label{secSpConcl}

Based on the spectral form of mathematical description of linear systems, a new technique for computer modeling linear filters is proposed in this paper. As examples, the Butterworth, Linkwitz--Riley, and Chebyshev filters of different orders are considered in detail. For each filter, the two-dimensional non-stationary transfer function corresponding to the transfer function is obtained.

In the proposed technique, input and output signals are represented by non-stationary spectral characteristics formed by expansion coefficients of these signals relative to the basis of the space of square-integrable functions. Algebraic operations are utilized to transform non-stationary spectral characteristics. For testing, cosines are chosen as the basis, but the basis can be arbitrary. In fact, the spectral method combines the advantages of analytical methods and the simple implementation inherent in numerical methods.

The advantage of the considered technique is that the output signal is modeled in continuous time. In addition to the above filters, it can be used for elliptic filters, as well as for the Bessel, Gaussian, and Legendre filters, etc. For these filters, the algorithm for deriving the transfer function is more complex. In particular, elliptic filters require the use of special functions (Jacobi elliptic functions), while the Gaussian filter is defined by an approximate representation of the Gaussian function as a partial sum of the Maclaurin series. However, once the transfer function is obtained, applying the spectral method for modeling these filters is no more difficult than for the Butterworth, Linkwitz--Riley, and Chebyshev filters.

\appendix

\section*{Appendix}

In Section~\ref{secSpectral}, the formulae are given for calculating elements of the matrix $S$ (for the pure time shift block by $\tau$) under the condition that basis functions are defined beyond the segment $\mathds{T}$ in a natural way. Further, we show that such a definition does indeed improve the quality of modeling linear filters in examples.

The usual definition of the two-dimensional non-stationary transfer function of the pure time shift block by $\tau$ assumes that basis functions $q(i,t)$ are extended by zero~\cite{SolSemPeshNed_79}: $q(i,t+\tau) = 0$ for $t+\tau \notin \mathds{T}$, $i = 0,1,2,\dots$ Consider this way in more detail, choosing cosines as the basis.

For the first case $\tau > 0$ ($S = S^+$),
\begin{gather*}
  S_{00}^+ = \frac{T-\tau}{T}, \ \ \ S_{0i}^+ = -\frac{\sqrt{2}}{i \pi} \sin \frac{i \pi \tau}{T}, \ \ \ S_{i0}^+ = \frac{\sqrt{2}}{i \pi} \sin \frac{i \pi (T-\tau)}{T}, \ \ \ S_{ii}^+ = \frac{T-\tau}{T} \cos \frac{i \pi \tau}{T} - \frac{1}{i \pi} \sin \frac{i \pi \tau}{T}, \\
  S_{ij}^+ = \frac{2}{(i^2 - j^2) \pi} \biggl( j \sin \frac{j \pi \tau}{T} - i (-1)^{i+j} \sin \frac{i \pi \tau}{T} \biggr),
\end{gather*}
and for the second case $\tau < 0$ ($S = S^-$),
\begin{gather*}
  S_{00}^- = \frac{T+\tau}{T}, \ \ \ S_{0i}^- = \frac{\sqrt{2}}{i \pi} \sin \frac{i \pi (T+\tau)}{T}, \ \ \ S_{i0}^- = \frac{\sqrt{2}}{i \pi} \sin \frac{i \pi \tau}{T}, \ \ \ S_{ii}^- = \frac{T+\tau}{T} \cos \frac{i \pi \tau}{T} + \frac{1}{i \pi} \sin \frac{i \pi \tau}{T}, \\
  S_{ij}^- = \frac{2}{(i^2 - j^2) \pi} \biggl( i \sin \frac{i \pi \tau}{T} - j (-1)^{i+j} \sin \frac{j \pi \tau}{T} \biggr),
\end{gather*}
where the simplest integration rules, as well as trigonometric identities are used to derive these formulae, and the indices take the following values: $i,j = 1,2,\dots$ and $i \neq j$.

Next, we consider how the redefinition of the matrix $S$ affects the error $\mathcal{E}$. In this case, we restrict ourselves to the Butterworth filters only since the conclusions for other filters are similar. The calculation results are given in Table~\ref{tabButterworthResult1x} for deterministic noise and in Table~\ref{tabButterworthResult2x} for random noise. The comparison of Tables~\ref{tabButterworthResult1} and~\ref{tabButterworthResult1x}, as well as~\ref{tabButterworthResult2} and~\ref{tabButterworthResult2x} shows that the filter quality improves by increasing the filter order when the basis functions are defined beyond the segment $\mathds{T}$ in a natural way.

In particular, if we extend basis functions by zero, then the function $x^*(t) = x(t+\tau_\varphi)$ will most likely be discontinuous at $t = T-\tau_\varphi$. The approximation of discontinuous functions is inferior in accuracy compared to the approximation of continuous and, especially, smooth functions~\cite{CanHusQuaZan_06}.

\ifnum \showtables = 1

\begin{table}[ht]
\centering
\renewcommand{\arraystretch}{1.1}

\caption{The error $\mathcal{E}$ for the Butterworth filters (deterministic noise, another definition of $S$)}\label{tabButterworthResult1x}
\begin{tabular}{ccccc}
  \hline
  $~~n~~$ & $L = 128$ & $L = 256$ & $L = 512$ & $L = 1024$ \\
  \hline
  2 & 0.048311 & 0.047964 & 0.047843 & 0.047794 \\
  3 & 0.026638 & 0.025594 & 0.025049 & 0.024766 \\
  4 & 0.019984 & 0.017576 & 0.015995 & 0.015353 \\
  5 & 0.020383 & 0.015491 & 0.013464 & 0.012332 \\
  6 & 0.019329 & 0.015356 & 0.013017 & 0.011682 \\
  \hline
\end{tabular}

\vskip 2ex

\caption{The error $\mathcal{E}$ for the Butterworth filters (random noise, another definition of $S$)}\label{tabButterworthResult2x}
\begin{tabular}{cccc}
  \hline
  $~~n~~$ & $L = 128$ & $L = 256$ & $L = 512$ \\
  \hline
  2 & 0.066455~(0.006000) & 0.066301~(0.006067) & 0.066147~(0.006097) \\
  3 & 0.064922~(0.006329) & 0.064288~(0.006304) & 0.064137~(0.006397) \\
  4 & 0.064450~(0.006415) & 0.063860~(0.006544) & 0.063458~(0.006520) \\
  5 & 0.065015~(0.006391) & 0.063500~(0.006562) & 0.063186~(0.006524) \\
  6 & 0.064580~(0.006484) & 0.063398~(0.006572) & 0.063071~(0.006603) \\
  \hline
\end{tabular}
\end{table}

\fi

The results for the third-order Butterworth filter are presented in graphical form in Figures~\ref{picButterworthResult1x} and~\ref{picButterworthResult2x}.

\ifnum \showfigures = 1

\begin{figure}[ht]
  \centering
  \includegraphics[scale = 0.85]{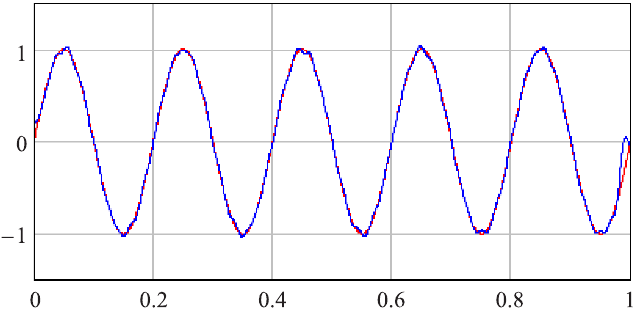}
  \caption{The graphs of the useful signal and the output signal of the Butterworth filter ($n = 3$, $L = 128$, deterministic noise, another definition of $S$)}\label{picButterworthResult1x}
  \vskip 2ex
  \includegraphics[scale = 0.85]{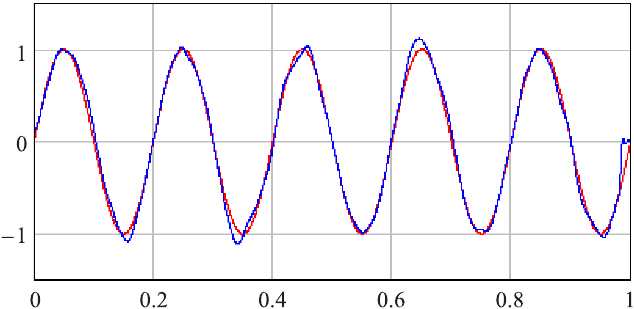}
  \caption{The graphs of the useful signal and the output signal of the Butterworth filter ($n = 3$, $L = 128$, random noise, another definition of $S$)}\label{picButterworthResult2x}
\end{figure}

\fi


\begin{thebibliography}{99}

\bibitem{Lam_82}
Lam, H.Y-E. {\em Analog and Digital Filters: Design and Realization}; Prentice Hall: Englewood Cliffs, NJ, USA, 1979.

\bibitem{Paa_03}
Paarmann, L.D. {\em Design and Analysis of Analog Filters: A Signal Processing Perspective}; Kluwer Academic Publ.: New York, NY, USA, 2003.

\bibitem{LanSotThe_Sensors23}
Langhammer, L.; Sotner, R.; Theumer, R. Various-order low-pass filter with the electronic change of its approximation. {\em Sensors} {\bf 2023}, {\em 23}(19), 8057. % 10.3390/s23198057

\bibitem{Fos_BSTJ24}
Foster, R.M. A reactance theorem. {\em The Bell System Technical Journal} {\bf 1924}, {\em 3}(2), 259--267.

\bibitem{Cau_31}
Cauer, W. {\em Siebschaltungen}; VDI-Verlag: Berlin, Germany, 1931.

\bibitem{But_EWWE30}
Butterworth, S. On the theory of filter amplifiers. {\em Experimental Wireless and the Wireless Engineer} {\bf 1930}, {\em 7}, 536--541.

\bibitem{Bode_JMP34}
Bode, H.W. A general theory of electric wave filters. {\em J. Math. Phys.} {\bf 1934}, {\em 13}(1--4), 275--362.

\bibitem{Dar_JMP39}
Darlington, S. Synthesis of reactance 4-poles which produce prescribed insertion loss characteristics. {\em J. Math. Phys.} {\bf 1939}, {\em 18}(1--4), 257--355.

\bibitem{Bel_PIRE62}
Belevitch, V. Summary of the history of circuit theory. {\em Proc. IRE} {\bf 1962}, {\em 50}(5), 848--855.

\bibitem{Dar_TCS84}
Darlington, S. A history of network synthesis and filter theory for circuits composed of resistors, inductors, and capacitors. {\em IEEE Trans. Circuits Syst.} {\bf 1984}, {\em 31}(1), 3--13.

\bibitem{Bal_21}
Bansal, D. PC-based data acquisition. In {\em Real-Time Data Acquisition in Human Physiology}; Academic Press, 2021; pp. 21--55.

\bibitem{GraMeyHer_SR25}
Graf, S.; Meyrand, P.; Herry, C.; Bem, T.; Tsai, F.-S. Self-supervised learning reduces label noise in sharp wave ripple classification. {\em Sci. Rep.} {\bf 2025}, {\em 15}, 7647. % 10.1038/s41598-025-90380-x

\bibitem{Pal_19}
Pal, S. ECG monitoring: Present status and future trend. In {\em Encyclopedia of Biomedical Engineering}; R. Narayan, ed.; Elsevier, 2019; Volume 3, pp. 363--379.

\bibitem{ChoBanKimPal_22}
Choudhury, A.D.; Banerjee, R.; Kimbahune, S.; Pal, A. Sensor signal analytics. In {\em New Frontiers of Cardiovascular Screening Using Unobtrusive Sensors, AI, and IoT}; Academic Press, 2022; pp. 61--89.

\bibitem{WasTanHaf_JRESTI24}
Waskita, N.I.; Tandungan, H.M.; Hafizh, R.; Suwaendi, S.J.; Magfirawaty, M. ESP32 and MAX30100 with Chebyshev filter for enhanced heart and oxygen measurement. {\em J. RESTI (Rekayasa Sistem dan Teknologi Informasi)} {\bf 2024}, {\em 8}(5), 651--657. % 10.29207/resti.v8i5.5945

\bibitem{HuaHeYangLinOuJia_Drones24}
Huang, J.; He, W.; Yang, D.; Lin, J.; Ou, Y.; Jiang, R.; Zhou, Z. Quantity monitor based on differential weighing sensors for storage tank of agricultural UAV. {\em Drones} {\bf 2024}, {\em 8}(3), 92. % 10.3390/drones8030092

\bibitem{MaiBouGab_JAIHC23}
Maitre, J.; Bouchard, K.; Gaboury, S. Data filtering and deep learning for enhanced human activity recognition from UWB radars. {\em J. Ambient Intell. Human. Comput.} {\bf 2023}, {\em 14}(6), 7845--7856. % 10.1007/s12652-023-04596-8

\bibitem{MudAzmAla_SR25}
Al Mudawi, N.; Azmat, U.; Alazeb, A.; Alhasson, H.F.; Alabdullah, B.; Rahman, H.; Liu, H.; Jalal, A. IoT powered RNN for improved human activity recognition with enhanced localization and classification. {\em Sci. Rep.} {\bf 2025}, {\em 15}, 10328. % 10.1038/s41598-025-94689-5

\bibitem{SungKangShim_JEET21}
Sung, S.; Kang, H.J.; Shim, H.; Shin, K.-H.; Cho, J.-Y. Compensation technique for delay times of various feedback filters in a three-phase control system for synchronous machines. {\em J. Electr. Eng. Technol.} {\bf 2021}, {\em 16}(6), 3069--3080. % 10.1007/s42835-021-00831-4

\bibitem{SriPuhiSib_JS23}
Sriyanto, S.P.D.; Puhi, A.R.; Sibuea, C.H.G. The performance of Butterworth and Wiener filter for earthquake signal enhancement: a comparative study. {\em J. Seismol.} {\bf 2023}, {\em 27}(1), 219--232.

\bibitem{RodValCas_Infra24}
Rodr\'{\i}guez-Qui\~{n}onez, J.C.; Valdez-Rodr\'{\i}guez, J.A.; Castro-Toscano, M.J.; Flores-Fuentes, W.; Sergiyenko, O. Inertial methodology for the monitoring of structures in motion caused by seismic vibrations. {\em Infrastructures} {\bf 2024}, {\em 9}(7), 116. % 10.3390/infrastructures9070116

\bibitem{AstZhiDem_20}
Astafiev, A.V.; Zhiznyakov, A.L.; Demidov, A.A. The use of Butterworth filter to compensate for noise in signals from Bluetooth low energy beacons in autonomous navigation systems. In Proceedings of the Int. Russ. Automation Conf. (RusAutoCon), Sochi, Russia, 2020; pp. 1117--1121. % 10.1109/RusAutoCon49822.2020.9208205

\bibitem{TianLiLiuLi_Sensors25}
Tian, C.; Li, F.; Liu, X.; Li, J. Optimized identity authentication via channel state information for two-factor user verification in information systems. {\em Sensors} {\bf 2025}, {\em 25}(8), 2465. % 10.3390/s25082465

\bibitem{Lin_JAES76}
Linkwitz, S.H. Active crossover networks for noncoincident drivers. {\em J. Audio Eng. Soc.} {\bf 1976}, {\em 24}(1), 2--8.

\bibitem{HarVenChenMutDick_IEEE13}
Harris, F.; Venosa, E.; Chen, X.; Muthyala, P.; Dick, C. An extension of the Linkwitz--Riley crossover filters for audio systems and their sampled data implementation. In Proceedings of the 20th Int. Conf. on Systems, Signals and Image Processing (IWSSIP), Bucharest, Romania, 2013; pp. 175--178. % https://doi.org/0.1109/IWSSIP.2013.6623482

\bibitem{CecBruNobTerVal_JAES23}
Cecchi, S.; Bruschi, V.; Nobili, S.; Terenziand, A.; V\"alim\"aki, V. Crossover networks: A review. {\em J. Audio Eng. Soc.} {\bf 2023}, {\em 71}(9), 526--551. % 10.17743/jaes.2022.0100

\bibitem{LutTosEva_01}
Lutovac, M.D.; To\v{s}i\'c, D.V.; Evans, B.L. {\em Filter Design for Signal Processing Using MATLAB and Mathematica}; Prentice Hall: Upper Saddle River, NJ, USA, 2001.

\bibitem{OppSch_14}
Oppenheim, A.V.; Schafer, R.W. {\em Discrete-Time Signal Processing}; Pearson Education Limited: Harlow, UK, 2014.

\bibitem{IngPro_17}
Ingle, V.K.; Proakis, J.G. {\em Digital Signal Processing Using MATLAB}; Cengage Learning: Stamford, CT, USA, 2017.

\bibitem{GuGaoLiuMaoLia_IEEE23}
Gu, G.; Gao, Z.; Liu, L.; Mao, W.; Liang, J. Design and simulation analysis of Bessel digital low-pass filters. In Proceedings of the 16th IEEE Int. Conf. on Electronic Measurement \& Instruments (ICEMI), Harbin, China, 2023; pp. 457--462. % 10.1109/ICEMI59194.2023.10270775

\bibitem{SolSemPeshNed_79}
Solodownikow, W.W.; Semjonow, W.W.; Peschel, M.; Nedo, D. {\em Berechnung von Regelsystemen auf Digitalrechnern: Anwendung von Spektral- und Interpolationsmethoden}; Verlag Technik: Berlin, Germany, 1979.

\bibitem{RybSot_TAC07}
Rybakov, K.A.; Sotskova, I.L. Spectral method for analysis of switching diffusions. {\em IEEE Trans. Autom. Control} {\bf 2007}, {\em 52}, 1320--1325. % 10.1109/TAC.2007.900841

\bibitem{Ryb_Comp25}
Rybakov, K. Spectral representation and simulation of fractional Brownian motion. {\em Computation} {\bf 2025}, {\em 13}, 19. % 10.3390/computation13010019

\bibitem{KornKorn_00}
Korn, G.A.; Korn, T.M. {\em Mathematical Handbook for Scientists and Engineers}; Dover Publ.: New York, NY, USA, 2000.

\bibitem{Weg_12}
Wegert, E. {\em Visual Complex Functions. An Introduction with Phase Portraits}; Birkh\"auser: Basel, Switzerland, 2012.

\bibitem{Bal_80}
Balakrishnan, A.V. {\em Applied Functional Analysis}; Springer: New York, NY, USA, 1981.

\bibitem{BagMikPanRyb_Springer20}
Baghdasaryan, G.Y.; Mikilyan, M.A.; Panteleev, A.V.; Rybakov, K.A. Spectral method for analysis of diffusions and jump diffusions. In {\em Smart Innovation, Systems and Technologies}; Springer: Singapore, 2020; Volume 173, pp. 293--314. % 10.1007/978-981-15-2600-8_21

\bibitem{RybYus_IOP20}
Rybakov, K.; Yushchenko, A. Spectral method for solving linear Caputo fractional stochastic differential equations. {\em IOP Conf. Ser. Mater. Sci. Eng.} {\bf 2020}, {\em 927}, 012077. % 10.1088/1757-899X/927/1/012077

\bibitem{Ryb_IOP21}
Rybakov, K. Modified spectral method for optimal estimation in linear continuous-time stochastic systems {\em J. Phys. Conf. Ser.} {\bf 2021}, {\em 1864}, 012025. % 10.1088/1742-6596/1864/1/012025

\bibitem{CanHusQuaZan_06}
Canuto, C.; Hussaini, M.Y.; Quarteroni, A.; Zang, T.A. {\em Spectral Methods. Fundamentals in Single Domains}; Springer: Berlin/Heidelberg, Germany, 2006.

\end{thebibliography}
\end{document}